\documentclass[pdflatex,sn-mathphys-num]{sn-jnl}

\usepackage{graphicx}%
\usepackage{multirow}%
\usepackage{amsmath,amssymb,amsfonts}%
\usepackage{amsthm}%
\usepackage{mathrsfs}%
\usepackage[title]{appendix}%
\usepackage{xcolor}%
\usepackage{textcomp}%
\usepackage{manyfoot}%
\usepackage{booktabs}%
\usepackage{algorithm}%
\usepackage{algorithmicx}%
\usepackage{algpseudocode}%
\usepackage{listings}%

\usepackage{txfonts}
\usepackage{tablefootnote}
\usepackage{longtable}
\usepackage{lscape}
\usepackage{booktabs}
\usepackage[utf8]{inputenc}
\usepackage{textcomp}
\usepackage{newunicodechar}
\newunicodechar{−}{\textminus}

\usepackage{hyperref}
\hypersetup{
    colorlinks = true,
    citecolor ={blue}
}

\usepackage{aas_macros}

\usepackage[T1]{fontenc}      
\usepackage[utf8]{inputenc}   
\usepackage{textcomp}         

\DeclareUnicodeCharacter{0306}{\u}  
\DeclareUnicodeCharacter{00E6}{\ae}  
\DeclareUnicodeCharacter{015B}{\'s}  
\DeclareUnicodeCharacter{0105}{\k a}  
\DeclareUnicodeCharacter{0119}{\k e}  
\DeclareUnicodeCharacter{0107}{\'c}   
\DeclareUnicodeCharacter{017C}{\.z}   
\DeclareUnicodeCharacter{0104}{\k A}  
\DeclareUnicodeCharacter{0118}{\k E}  
\DeclareUnicodeCharacter{0106}{\'C}   
\DeclareUnicodeCharacter{017B}{\.Z}   
\DeclareUnicodeCharacter{0161}{\v s}  
\DeclareUnicodeCharacter{010D}{\v c}  

\begin{document}

\title[Article Title]{Observing bright pulsating white dwarfs with PLATO: A new window into the late stages of stellar evolution}


\author*[1]{\fnm{Murat} \sur{Uzundag}}\email{muratuzundag.astro@gmail.com}

\author[2,3]{\fnm{Alejandro H.} \sur{Córsico}}

\author[4]{\fnm{Nicholas} \sur{Jannsen}}

\author[5]{\fnm{Mukremin} \sur{Kilic}}

\author[6]{\fnm{Pierre} \sur{Bergeron}}

\author[2,3]{\fnm{Leandro G.} \sur{Althaus}}

\author[7]{\fnm{J. J.} \sur{Hermes}}

\author[8]{\fnm{Ingrid} \sur{Pelisoli}}

\author[9]{\fnm{Keaton J.} \sur{Bell}}

\author[2,3]{\fnm{Francisco C.} \sur{De Gerónimo}}

\author[2,3]{\fnm{Leila M.} \sur{Calcaferro}}

\author[10]{\fnm{Zs\'ofia} \sur{Bogn\'ar}}

\author[11]{\fnm{Val\'erie Van} \sur{Grootel}}

\author[12]{\fnm{María E.} \sur{Camisassa}}

\author[13,14]{\fnm{Paulina} \sur{Sowicka}}

\author[15]{\fnm{Steven D.} \sur{Kawaler}}

\author[16]{\fnm{S. O.} \sur{Kepler}}

\author[17]{\fnm{Roberto} \sur{Silvotti}}

\author[2,3]{\fnm{Marcelo M.} \sur{Miller Bertolami}}

\author[18]{\fnm{Margarida} \sur{Cunha}}

\affil*[1]{\orgdiv{Institute of Astronomy}, \orgname{KU Leuven}, 
\orgaddress{\street{Celestijnenlaan 200D}, \city{Leuven}, \postcode{3001}, \country{Belgium}}}

\affil[2]{\orgdiv{Grupo de Evolución Estelar y Pulsaciones, Facultad de Ciencias Astronómicas y Geofísicas}, 
\orgname{Universidad Nacional de La Plata}, 
\orgaddress{\street{Paseo del Bosque s/n}, \city{La Plata}, \postcode{1900}, \country{Argentina}}}

\affil[3]{\orgdiv{Instituto de Astrofísica de La Plata (IALP), CCT La Plata}, 
\orgname{CONICET-UNLP}, \country{Argentina}}

\affil[4]{\orgdiv{Isaac Newton Group of Telescopes}, 
\orgaddress{
\street{Apartado de correos 321, 
\city{ Santa Cruz de La Palma}, 
\state{Canary Islands}, 
\postcode{E-38700}, 
\country{Spain}}}}

\affil[5]{\orgdiv{Homer L. Dodge Department of Physics and Astronomy}, 
\orgname{University of Oklahoma}, 
\orgaddress{\street{440 W. Brooks St.}, \city{Norman}, \state{OK}, \postcode{73019}, \country{USA}}}

\affil[6]{\orgdiv{Département de Physique}, 
\orgname{Université de Montréal}, 
\orgaddress{\street{C.P. 6128, Succ. Centre-Ville}, \city{Montréal}, \state{QC}, \postcode{H3C 3J7}, \country{Canada}}}

\affil[7]{\orgdiv{Department of Astronomy}, \orgname{Boston University}, 
\orgaddress{\street{725 Commonwealth Avenue}, \city{Boston}, \state{MA}, \postcode{02215}, \country{USA}}}

\affil[8]{\orgdiv{Department of Physics}, \orgname{University of Warwick}, 
\orgaddress{\street{Gibbet Hill Road}, \city{Coventry}, \postcode{CV4 7AL}, \country{UK}}}

\affil[9]{\orgdiv{Department of Physics}, \orgname{CUNY Queens College}, 
\orgaddress{\street{65-30 Kissena Blvd}, \city{Flushing}, \state{NY}, \postcode{11367}, \country{USA}}}

\affil[10]{\orgname{HUN-REN Research Centre for Astronomy and Earth Sciences, Konkoly Observatory, MTA Centre of Excellence}, 
\orgaddress{\street{Konkoly Thege Mikl\'os \'ut 15-17}, \city{Budapest}, \postcode{H-1121}, \country{Hungary}}}

\affil[11]{\orgdiv{Space Sciences, Technologies and Astrophysics Research (STAR) Institute}, 
\orgname{Université de Liège}, 
\orgaddress{\street{19C Allée du 6 Août}, \city{Liège}, \postcode{B-4000}, \country{Belgium}}}

\affil[12]{\orgdiv{Departament de Física}, 
\orgname{Universitat Politècnica de Catalunya}, 
\orgaddress{\street{c/Esteve Terrades 5}, \city{Castelldefels}, \postcode{08860}, \country{Spain}}}

\affil[13]{\orgname{Instituto de Astrofísica de Canarias}, 
\orgaddress{\city{La Laguna}, \postcode{E-38205}, \country{Spain}}}

\affil[14]{\orgdiv{Departamento de Astrofísica}, 
\orgname{Universidad de La Laguna}, 
\orgaddress{\city{La Laguna}, \postcode{E-38206}, \country{Spain}}}

\affil[15]{\orgdiv{Department of Physics and Astronomy}, 
\orgname{Iowa State University}, 
\orgaddress{\city{Ames}, \state{IA}, \postcode{50011}, \country{USA}}}

\affil[16]{\orgdiv{Instituto de Física}, 
\orgname{Universidade Federal do Rio Grande do Sul}, 
\orgaddress{\city{Porto Alegre}, \postcode{91501-970}, \state{RS}, \country{Brazil}}}

\affil[17]{\orgname{INAF–Osservatorio Astrofisico di Torino}, 
\orgaddress{\street{Strada Osservatorio 20}, \city{Pino Torinese}, \postcode{10025}, \country{Italy}}}

\affil[18]{\orgdiv{Instituto de Astrofísica e Ciências do Espaço}, 
\orgname{Universidade do Porto, CAUP}, 
\orgaddress{\street{Rua das Estrelas}, \postcode{PT4150-762}, \city{Porto}, \country{Portugal}}}


\abstract{
We present the scientific case for exploiting the capabilities of the PLATO mission to study bright pulsating white dwarfs across a wide spectral range, including hydrogen-deficient types (GW Vir and DBV stars) and hydrogen-rich classes (classical DAVs, pulsating extremely low-mass DA white dwarfs, and ultra-massive DA white dwarfs).
PLATO’s exceptional photometric precision, long-duration continuous monitoring, and extensive sky coverage promise transformative advances in white dwarf asteroseismology. Our key objectives include probing the internal structure and chemical stratification of white dwarfs, detecting secular changes in pulsation modes over extended timescales, and discovering rare or previously unknown classes of pulsators.
To assess feasibility, we constructed a sample of 650 white dwarf candidates ($G \leq 17$) identified within PLATO’s Southern LOPS2 field using the PLATO complementary science catalogue combined with \textit{Gaia} DR3, and derived atmospheric parameters through photometric modeling. This sample comprises 118 DA white dwarfs (including 23 ZZ Ceti candidates), and 41 non-DAs (including 35 DBV candidates). Simulated observations with {\tt PlatoSim} demonstrate that PLATO will detect white dwarf pulsation modes with amplitudes as low as $\sim$0.1\,mma, depending on stellar magnitude, observation duration, pixel location, and the number of contributing cameras. We provide detailed detection limits and visibility forecasts for known pulsators across a representative range of these parameters.
Furthermore, we emphasize strong synergies with \textit{Gaia} astrometry, TESS photometry, and targeted spectroscopic campaigns, which together will enable robust mode identification and detailed stellar modeling. Collectively, these efforts will unlock unprecedented insights into white dwarf origins, evolution and internal physics, and the fate of their planetary systems.}




\maketitle

\section{Introduction}\label{intro}

White dwarfs (WDs) are the ultimate fate of stars with initial masses below roughly $7$–$11\,M_{\odot}$, encompassing over 95\% of all stars in the Milky Way, including our own Sun \citep{2001PASP..113..409F,2015ApJ...810...34W,2010A&ARv..18..471A,2022PhR...988....1S}. As such, they offer unique insights into the formation, evolution, and final stages of both stellar and planetary systems. A subset of WDs exhibit photometric variability due to various phenomena, including stellar pulsations, binarity, and transiting planetary debris \citep[][]{2021ApJ...912..125G,2024ApJ...967..166S}. Stellar pulsations are particularly effective tools for probing the interiors of WDs, which undergo at least one phase of pulsational instability during their evolution. Indeed, pulsating WDs allow for asteroseismology —the study of internal structure through stellar oscillations— which has proven instrumental in refining our understanding of stellar physics \citep{2008ARA&A..46..157W,2008PASP..120.1043F,2019A&ARv..27....7C}.

WD pulsations are global, non-radial $g$ (gravity)-mode oscillations with harmonic degree $\ell= 1, 2$ and periods ranging from $\sim 100$ seconds to $\sim 7000$ seconds and amplitudes typically below 0.4 mag \citep{2008ARA&A..46..157W, 2019A&ARv..27....7C}. These oscillations are excited by different mechanisms, depending on the atmospheric composition and evolutionary stage of the star, related to partial ionization zones \citep[$\kappa-\gamma$ mechanism;][]{1981A&A...102..375D,1981A&A....97...16D, 1982ApJ...252L..65W, 1983ApJ...268L..27S, 1984ApJ...281..800S} and convection \citep["convective driving" mechanism;][]{1991MNRAS.251..673B, 1999ApJ...511..904G, 2018ApJ...863...82L} in the outer layers. 

Pulsations in WDs are extremely sensitive to the internal chemical stratification of the star, allowing for detailed asteroseismic modeling. By comparing the observed pulsation periods and period spacings with those predicted by theoretical models, we can infer key structural parameters such as effective temperature ($T_{\rm eff}$), surface gravity (log $g$), and total stellar mass ($M_{\star}$) \citep[e.g.,][]{2012MNRAS.420.1462R,2014ApJ...794...39B, 2018Natur.554...73G,2023MNRAS.526.2846U}. Even more, asteroseismology enables us to probe the chemical composition of the stellar interior—including the relative abundances of oxygen (O) and carbon (C) \citep{2003ApJ...587L..43M,2022ApJ...935...21C,2023ApJ...954...51C} —as well as to estimate the extent of core crystallization \citep{2025arXiv250517177C} and the thickness of the surface hydrogen (H) or helium (He) layers \citep{2017A&A...599A..21D,2018A&A...613A..46D} by studying the effects of "mode trapping". Mode trapping in pulsating WDs refers to a phenomenon where certain pulsation modes are partially trapped or confined to specific regions within the star due to its internal chemical structure \citep{1981ApJ...245L..33W,1992ApJS...80..369B,1996ApJ...468..350B}. In addition, stellar rotation and magnetic fields can be investigated by examining frequency splitting in the Fourier spectra \citep{2017ApJS..232...23H, 2025ApJ...981...72R}. Also, pulsating WDs allow us to investigate the properties of fundamental particles like axions and neutrinos and the possible variation of fundamental constants \citep{2012MNRAS.424.2792C,2013JCAP...06..032C}. Finally, pulsating massive WDs can potentially be used to study the impact of General Relativity on white dwarf structure and pulsations \citep{2022A&A...668A..58A,2023MNRAS.523.4492A,2023MNRAS.524.5929C}.


The majority of spectroscopically identified WDs possess H-rich (DA) atmospheres, with estimates generally indicating that 70–80\% of WDs are H-dominated.
This percentage varies with effective temperature, as shown in Figure 2 of \citet{2024Ap&SS.369...43B}, and is consistent with several spectroscopic studies, including \citet{K2019}, \citet{2024MNRAS.527.8687O}, and \citet{kilic25}.
Pulsating DA WDs —also known as ZZ Ceti stars and DAVs—represent the most numerous class of pulsating WDs. These stars exhibit non-radial $g$-mode pulsations within a narrow instability strip spanning effective temperatures of $T_{\rm eff} \sim 10\,500 - 13\,000$\,K and surface gravities in the range $7.5 < \log g < 9.2$. High-precision photometric observations from space missions such as \textit{Kepler}/K2 \citep{2010Sci...327..977B,2014PASP..126..398H} and TESS \citep{2015JATIS...1a4003R} have been revolutionary in the study of DAVs. Indeed, the \textit{Kepler}/K2 satellite has allowed the discovery of new DAVs, the derivation of their rotation rates, and the unexpected finding of outbursting ZZ Cetis near the red edge of the DAV instability strip, as well as the discovery of a clear dichotomy of oscillation mode-line widths in the power spectrum \citep{2014MNRAS.438.3086G, 2015ASPC..493..169G, 2016MNRAS.457.2855G, 2015ApJ...809...14B, 2016ApJ...829...82B, 2017ApJ...851...24B, 2017ApJS..232...23H, 2017PhDT........14C}. 
The TESS mission, on the other hand, has led to observations of numerous already known DAVs \citep{2020A&A...638A..82B,2023A&A...674A.204B} as well as the discovery of many new stars of this type, expanding the number of confirmed DAVs to over 500 objects \citep{2022MNRAS.511.1574R, 2023MNRAS.518.1448R, 2023MNRAS.526.2846U, 2025ApJ...984..112R}, and establishing them as key targets for asteroseismological investigations of WDs. TESS mission has also allowed the derivation of the rotation periods of many ZZ Ceti stars \citep{2024A&A...684A..76B}. Among the H-rich pulsating WDs, we also find ELMVs, which are He-core extremely low-mass (ELM) variable WDs \citep{hermes12,kilic15}. These objects are the products of binary evolution, having lost most of their mass through interaction with a companion, and therefore never ignited He in their cores. Asteroseismology of ELMVs provides a unique window into their internal structure and evolutionary history, offering key constraints on binary mass-loss processes and the physics of compact binary systems \citep{2013A&A...557A..19A, 2014A&A...569A.106C, 2014ApJ...793L..17C, 2016A&A...585A...1C, 2017A&A...600A..73C, 2017A&A...607A..33C, 2018A&A...620A.196C, 2025A&A...699A.280A}.
TESS observations have allowed the discovery and modeling of pulsations of the ELMV star GD~278 \citep{2021ApJ...922..220L}, and also the study of its internal rotation \citep{2023A&A...673A.135C}. Finally, among H-rich WD pulsators, there are the "hot DAVs", which are DA WDs that show photometric variability at $T_{\rm eff} \sim 30\,000$ K \citep{2013MNRAS.432.1632K, 2020MNRAS.497L..24R}, whose existence was anticipated by the theoretical calculations of \cite{2005EAS....17..143S,2007AIPC..948...35S}.
To this list, we must add the hypothetical "warm" ($T_{\rm eff} \sim 19\,000$ K) pulsating DA WDs,  hotter than ZZ Ceti stars, which were predicted to exist in theoretical studies more than 40 yr ago \citep{1982ApJ...252L..65W}, but that have not been detected yet \citep{2020A&A...633A..20A}.


The remaining $\sim$20-30\% of WDs are characterized by H-deficient atmospheres, encompassing spectral types such as PG~1159, DO, DB, and DQ \citep{Koester2015, 2024Ap&SS.369...43B}. These stars represent a more diverse and less well-understood population compared to their H-rich counterparts. Among the H-deficient pulsators, two main classes of pulsationally unstable WDs are observed: GW Vir stars (historically also referred to as the DOV stars; see Sec. 10 in \citealt{2023ApJS..269...32S} for discussion) —very hot and luminous, pre-WD objects descended from PG~1159 star progenitors— and the DBV (or V777 Her) stars, which are He-atmosphere pulsators with effective temperatures near $T_{\rm eff} \sim 25\,000$\,K. The evolutionary pathways that lead to H-deficient WDs remain a subject of ongoing investigation \citep{2019A&ARv..27....7C}. For example, DBVs may descend from PG~1159 stars after evolving through the DO phase, characterized by very hot He-rich atmospheres \citep{2010A&ARv..18..471A,2024Ap&SS.369...43B}. Alternatively, they may form through the merger of two WDs. H-deficient pulsators are intrinsically rarer than DAVs, and their instability strips—defined regions in the Hertzsprung-Russell (H-R) diagram where pulsations are expected—are less well established. In many cases, the boundaries predicted by theoretical models do not coincide with those determined from observations \citep{2008PASP..120.1043F,2019A&ARv..27....7C,2022ApJ...927..158V}, highlighting the need for further discoveries and detailed asteroseismic analyses to refine our understanding of these rare objects.
Regarding H-deficient pulsating WDs and pre-WDs such as GW Vir and DBV stars, space missions such as \textit{Kepler}/K2 and TESS have been highly productive, yielding discoveries of new pulsating stars, new periods in known pulsating stars, and confirmation of previously detected periods from ground-based observations \citep{2021A&A...645A.117C,2021A&A...655A..27U,2022A&A...659A..30C,2022MNRAS.513.2285U, 2022ApJ...936..187O,2022A&A...668A.161C, 2024A&A...686A.140C}. In many cases, it has been possible to infer the total
mass, internal chemical layering, and rotation rates using asteroseismic techniques such as asymptotic period spacing, period-to-period fits and rotational splitting. Also, it has been possible to infer the asteroseismic distances and compare them with the \textit{Gaia} distances. Another interesting aspect of seismological modeling of these stars is the prediction of the rate of change of their periods, which can be compared with the observed rates
\citep[for example, see][]{2022ApJ...936..187O}.

In this paper, we propose a coordinated effort to utilize data from upcoming space missions—particularly PLATO (PLAnetary Transits and Oscillations of stars) —along with ground-based follow-up to conduct asteroseismological analyses of pulsating WDs across all subtypes. Such efforts will significantly advance our understanding of stellar remnants and the fate of planetary systems.

\section{The Complementary Science Program}

To harness the full scientific capabilities of the PLATO mission beyond its core science objectives, the PLATO Complementary Science (CS) Program—also referred to as PLATO-CS—has been established \citep{2024eas..conf..871T, 2024A&A...692R...1A}. This initiative is designed to support diverse scientific investigations by allocating 8\% of PLATO’s telemetry budget to non-core science cases through a dedicated Guest Observer (GO) program \citep{2024eas..conf.2579H}. These open, competitive calls will allow the broader astronomical community to propose studies of specific astrophysical targets not covered by the core mission. PLATO-CS plays a pivotal role in shaping and preparing these GO calls, ensuring high-impact science opportunities are made available.
This 8\% telemetry allocation enables the monitoring of thousands to tens of thousands of strategically selected objects, including rare and compact stars such as WDs and hot subdwarfs. In this paper, we present the science cases we intend to pursue within the GO framework, highlighting the unique potential of PLATO data to advance our understanding of compact pulsators and related phenomena.

\section{Science Goals}

The interior properties of WDs encode the history of the star's prior evolution \citep{2006PASP..118..183W,2010A&ARv..18..471A, 2022PhR...988....1S}. WD asteroseismology not only reveals the current internal structure of WDs, but also constrains the physical processes that shaped their progenitors, such as core growth during He burning, thermal pulses on the asymptotic giant branch (AGB), and late thermal pulses \citep{2008ARA&A..46..157W,2008PASP..120.1043F,2019A&ARv..27....7C, 2024ApJ...964...30C}. For example, DBVs can help trace the evolution of He-rich post-AGB stars, while GW Vir stars may retain signatures of very late thermal pulse events. To fully exploit this potential, it is crucial to detect a large number of pulsation modes. The more modes we observe, the more accurately we can constrain the internal structure and test evolutionary models. However, many WDs observed so far have shown only a few dominant pulsation periods, often due to observational limitations such as cadence, duration, and noise level.

PLATO’s long-duration, high-precision, and high-cadence observations offer a transformative opportunity to expand the mode detection capabilities for WD asteroseismology. By enabling continuous, high signal-to-noise monitoring of bright pulsators, PLATO will allow us to:
\begin{itemize}
    \item Track temporal evolution of pulsation frequencies and amplitudes, probing processes such as cooling, mode trapping, and rotational splitting;
    \item Identify rotational multiplets to directly measure internal rotation rates;
    \item Search for new or rare classes of pulsators, including ultra-massive and extremely low mass WDs;
     \item Search for and identify pulsations in candidate objects such as hot DAVs, for which the origin of variability remains uncertain. Expanding the sample of confirmed pulsators in this category is crucial for refining pulsation theory and testing models of WD formation and evolution;
    \item Characterize the crystallization process in ultra-massive pulsating WDs through the detection of the period spacing.
\end{itemize}

These goals will be addressed through a coordinated GO program, focusing on a well-curated sample of known and candidate pulsators. PLATO’s long-baseline observations and high cadence will be essential for resolving low-amplitude and closely spaced pulsation modes. The goals will be further enhanced through synergies with \textit{Gaia} astrometry, TESS light curves, and ground-based spectroscopic follow-up, which together provide independent constraints on stellar parameters and mode identification.

In summary, PLATO will significantly extend the frontiers of WD asteroseismology, offering an unprecedented view into the physics of stellar remnants and the evolutionary history of low- and intermediate-mass stars.

\subsection{DAV Stars}

DAVs (ZZ Ceti stars) are the most common class of WD pulsators, typically found at $T_{\rm eff}\sim  12,000$ K. Despite their ubiquity, several open questions remain regarding their instability strip boundaries, internal structure, and evolutionary histories. With its high precision and extended time coverage, PLATO offers a transformative opportunity to advance DAV studies by:

\begin{itemize}
\item Expanding the sample of hot (blue-edge) DAVs, which are currently underrepresented, to refine instability strip boundaries and test driving mechanisms,
\item Searching for temporary overall brightness changes coupled with the pulsations, such as outbursts,
\item Discovering many new ZZ Ceti stars, which could potentially allow confirmation of the existence of a mass range of the H envelope in DA WDs, as past analyses have suggested,
\item Searching for elusive warm DAVs (T$_\mathrm{eff}$~$\sim$19,000 K) with very thin H layers, predicted to be unstable to $g$ modes, but yet to be observed,
\item Resolving rotational multiplets to probe internal rotation and angular momentum loss,
\item Detecting long-term amplitude and frequency modulations to investigate  mode coupling, convection, non-linear effects, and possible evolutionary changes,
\item Enabling detailed asteroseismology to constrain crystallization processes, core composition, and phase transitions,
\item Identifying and characterizing polluted DAVs to explore accretion geometry and link pulsations to external planetary debris.
\end{itemize}

\subsection{ELMV Stars}

ELMVs (pulsating low-mass and extremely low-mass WDs) are He-core WDs formed in compact binary systems after undergoing intense mass loss. They exhibit long-period $g$-mode pulsations that make them valuable asteroseismic targets. With PLATO, we aim to:
\begin{itemize}
    \item Detect new ELMV pulsators suitable for asteroseismology,
    \item Constrain H envelope thickness and mass-loss history,
    \item Measure surface rotation through resolved multiplets.
\end{itemize}

\subsection{Pulsating Ultra-massive DA WDs}

UM (ultra-massive) DA WDs are characterized by stellar masses $M_{\star} \gtrsim 1.05\,M_{\odot}$. Pulsating UM DA WDs constitute an excellent opportunity for studying a wide variety of fundamental processes, including the final stages of stellar evolution, the progenitors of type Ia supernovae, the outcomes of WD mergers, the physics of the super-asymptotic giant branch (SAGB), and the effects of core crystallization. These stars are also often associated with strong magnetic fields. Different evolutionary scenarios are proposed for the formation of UM WDs: (1) isolated evolution of intermediate-mass stars, for which C ignites in their interior, leading to a ONe-core UM WD, (2) isolated evolution for which the presence of weak winds or fast rotation could avoid C-ignition, resulting in a CO-core UM WDs, and (3) post-merger evolution of binary WDs, leading to ONe- and potentially CO-core compositions. The internal composition and origin of UM WDs remain uncertain, and pulsations offer one of the few direct methods (if not the only one) to probe their cores. With PLATO’s precision photometry, we aim to:

\begin{itemize}
    \item Identify and characterize new pulsating UM DAVs,
    \item Search for UM DAQV stars, like WD~J0551+4135\footnote{This is the only pulsating DAQ WD with a C and H atmosphere known \citep{2020NatAs...4..663H}}, which challenge WD formation theories;
    \item Measure period spacing patterns to test core composition and crystallization theory,
    \item Search for mode trapping and rotational splitting to constrain envelope layering and angular momentum,
    \item Distinguish between single-evolution and merger-origin scenarios through detailed asteroseismic modeling.
\end{itemize}



\subsection{GW Vir Stars}

GW Vir stars, also known as PG~1159 pulsators, are hot, hydrogen-deficient post-AGB stars in a brief but crucial phase of stellar evolution. The classification of GW Vir stars also includes the pulsating Wolf-Rayet central stars of planetary nebula (WC) and early-WC (WCE)
stars because they share the pulsation properties of pulsating
PG~1159 stars \citep{2007ApJS..171..219Q}. They exhibit non-radial $g$-mode pulsations that provide insight into their internal structure and evolutionary state. Despite being located within the theoretical instability strip, only about 36-50\% of these stars are observed to pulsate \citep{2021A&A...655A..27U,2022MNRAS.513.2285U,2023ApJS..269...32S}, indicating an incomplete understanding of the driving mechanisms and instability boundaries \citep{2007ApJS..171..219Q}. The prototypical and most studied object of this class is PG~1159-035 (GW Vir). Results from the most recent asteroseismological analyses of this star are disturbing, since the seismological model (which accurately reproduces the observed periods) lies outside the instability domain of these 
stars \citep{2008A&A...478..869C,2022ApJ...936..187O}. Also, the observed rates of period changes  \citep{2008A&A...489.1225C} are ten times smaller than those predicted by the seismological model. In addition, the rates of period change predicted by the best-fit model
are positive for all modes, and thus they do not agree with the observed positive and negative values. Reconciling theoretical predictions with observations remains a challenge.

PLATO's precise and long-duration photometry will significantly advance our knowledge of these stars by enabling the following:

\begin{itemize}
\item Expanding the sample size to better constrain the efficiency and boundaries of the GW Vir instability strip,
\item Searching for correlations between surface abundances and pulsation occurrence, in order to understand the strip’s impurity and test proposed excitation/damping mechanisms,
\item Monitoring secular period changes caused by rapid cooling and contraction. GW Vir stars evolve quickly, and PLATO's long-term coverage will enable us to determine these evolutionary changes on timescales of just a few years.
\end{itemize}


\subsection{DBV Stars}

DBV stars are He-atmosphere WDs pulsating at $T_{\rm eff}\sim 25,000$ K. Increasing the number of known DBVs is critical for constraining their poorly understood evolutionary origins, particularly the role of convective dilution versus alternative channels like binary mergers. PLATO's high-precision photometry and long-time baselines will enable us to:

\begin{itemize}
\item Expand the population of DBVs for asteroseismic modeling, to probe internal structure and test evolutionary scenarios,
\item Search for pulsating UM DBVs. Being much hotter and less crystallized than UM DAVs, if they exist, they would allow probing deeper regions with the $g$-mode pulsations, potentially providing information on the structure and chemical composition of the core,
\item Identify stable pulsators near the blue edge to measure WD cooling rates and constrain plasmon neutrino and axion emission from rates of period changes,
\item Monitor period changes and amplitude modulations to investigate nonlinear mode interactions and long-term evolution,
\item Empirically refine the location and extent of the DBV instability strip to calibrate convective efficiency and improve diffusion models relevant for planetary debris analysis,
\item Searching for outbursts in cool DBVs ---there are hints that outbursts could have been observed in at least one DBV in the past \citep[the "forte" or  "sforzando" event observed in GD~358;][]{2003A&A...401..639K,2005A&A...432..175C,2009ApJ...693..564P}.
\end{itemize}



\section{Sample selection}\label{sec:sample}

As part of a larger simulation study for the `Pulsating Stars' work package of the PLATO-CS, \cite{2025A&A...694A.185J} generated a PLATO asteroseismic mock catalogue (MOCKA). An important delivery of MOCKA is a \textit{Gaia} DR3 source catalogue of stars brighter than $G<17$ within the first Southern long-duration observational phase \citep[LOPS2;][]{nascimbeni2025plato}. With six co-pointing `normal' cameras (N-CAMs), situated in four camera groups each with an opening angle of $9.2^{\circ}$ relative to the platform pointing, each PLATO pointing field display a characteristic geometric feature. This is also known as the N-CAM visibility flower, and is a result of the partial overlapping number of N-CAMs, counting $n_{\rm CAM} \in \{6, 12, 18, 24\}$ (which will be shown in Sect.~\ref{sec:simulations}).

\begin{figure}
\centering
\includegraphics[width=1.0\columnwidth]{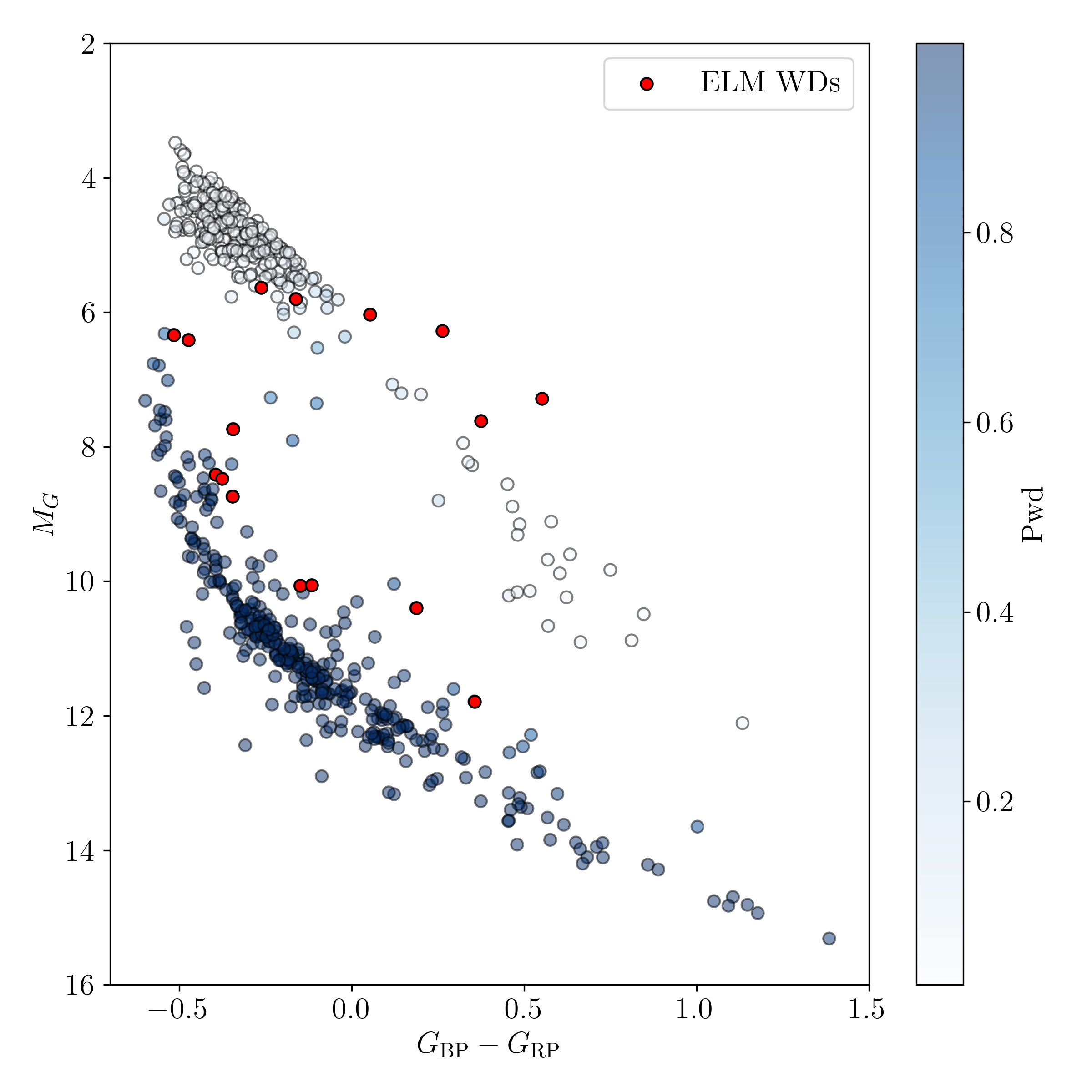}
\caption{H-R diagram of white dwarf candidates identified in the PLATO-CS input catalog \citep{2025A&A...694A.185J}. The color bar indicates the WD probability from \citet{2021MNRAS.508.3877G}. Extremely low-mass (ELM) WD candidates, selected from the PLATO-CS catalog and cross-matched with the sample of \citet{2019MNRAS.488.2892P}, are shown in red.
}
\label{fig:cmd_WD}
\end{figure}

The PLATO--LOPS2 \textit{Gaia} DR3 (from hereon PLATO-CS) catalog contains almost eight million sources in the magnitude range $2 \leq G \leq 17$.
To identify white dwarf (WD) candidates within this sample, we cross-matched PLATO-CS with the catalog from \citet{2021MNRAS.508.3877G} of high-confidence WDs based on \textit{Gaia} EDR3, as well as the catalog of ELM WD candidates from \citet{2019MNRAS.488.2892P}. Through this combined approach, we identified an initial sample of 649 WD candidates within the LOPS2 region.
Figure~\ref{fig:cmd_WD} shows the H-R diagram of our initial WD candidates identified in the PLATO-CS input catalog. Each candidate was assigned a WD probability following \citet{2021MNRAS.508.3877G}, with increasing probability the darker the shade of blue. 
      

\section{Photometric model atmosphere analysis }

To determine the fundamental atmospheric parameters of our white dwarf sample, we use the photometric method \citep{bergeron97,2019ApJ...876...67B},
which relies on multi-band photometry combined with precise astrometric data from \textit{Gaia} DR3.

We use broad-band photometry from Skymapper ($uvgriz$, available for 99\% of the sample) and Pan-STARRS ($grizy$, available for 23\% of the sample),
along with \textit{Gaia} parallaxes, to constrain both the effective temperature and the solid angle, $\pi (R/D)^2$, of each
star. Given the accurate distances from \textit{Gaia}, we directly estimate the stellar radii, which are then combined with white dwarf evolutionary
models to derive the stellar masses. For a small sample of stars, where the Skymapper or Pan-STARRS photometry is unavailable or contaminated by
a neighboring source, we instead rely on the \textit{Gaia} $G$, $G_{\rm BP}$, and $G_{\rm RP}$  photometry. 

The observed magnitudes are first converted into average fluxes, which are then compared to synthetic fluxes generated from model atmospheres corresponding to various temperatures and chemical compositions. We minimize the $\chi^2$ difference between observed and model fluxes across all bandpasses using the Levenberg–Marquardt nonlinear least-squares algorithm \citep{1986nras.book.....P}, yielding the best-fit atmospheric parameters. Uncertainties on the fitted parameters are derived from the covariance matrix of the fitting routine. For parameters calculated from these fits, such as mass and radius, uncertainties are propagated in quadrature from the input measurement errors \citep[see][for a detailed discussion of the photometric method and the derived errors]{bergeron97,2019ApJ...876...67B}.

Our analysis utilizes pure H and mixed H/He models with $\log$ (H/He) = $-5$ as described in \citet{2019ApJ...876...67B}. Only 79 of our
targets have spectra available in the Montreal White Dwarf Database \citep{dufour17}, though a larger number has spectral types provided
in the literature. For the ones without follow-up spectroscopy, we provide both pure H and mixed H/He model fits as representative solutions.
We adopt the evolutionary models of \citet{2020ApJ...901...93B}, which assume carbon-oxygen cores
with helium and hydrogen envelope layer masses of $q(\mathrm{He}) = 10^{-2}$, and $q(\mathrm{H}) = 10^{-4}$ (for H-atmosphere white dwarfs) or
$10^{-10}$ (for He-atmosphere white dwarfs). These assumptions are representative of DA and DB white dwarfs, respectively.

\begin{figure}
\center
\includegraphics[width=\columnwidth,height=0.8\textheight,keepaspectratio]{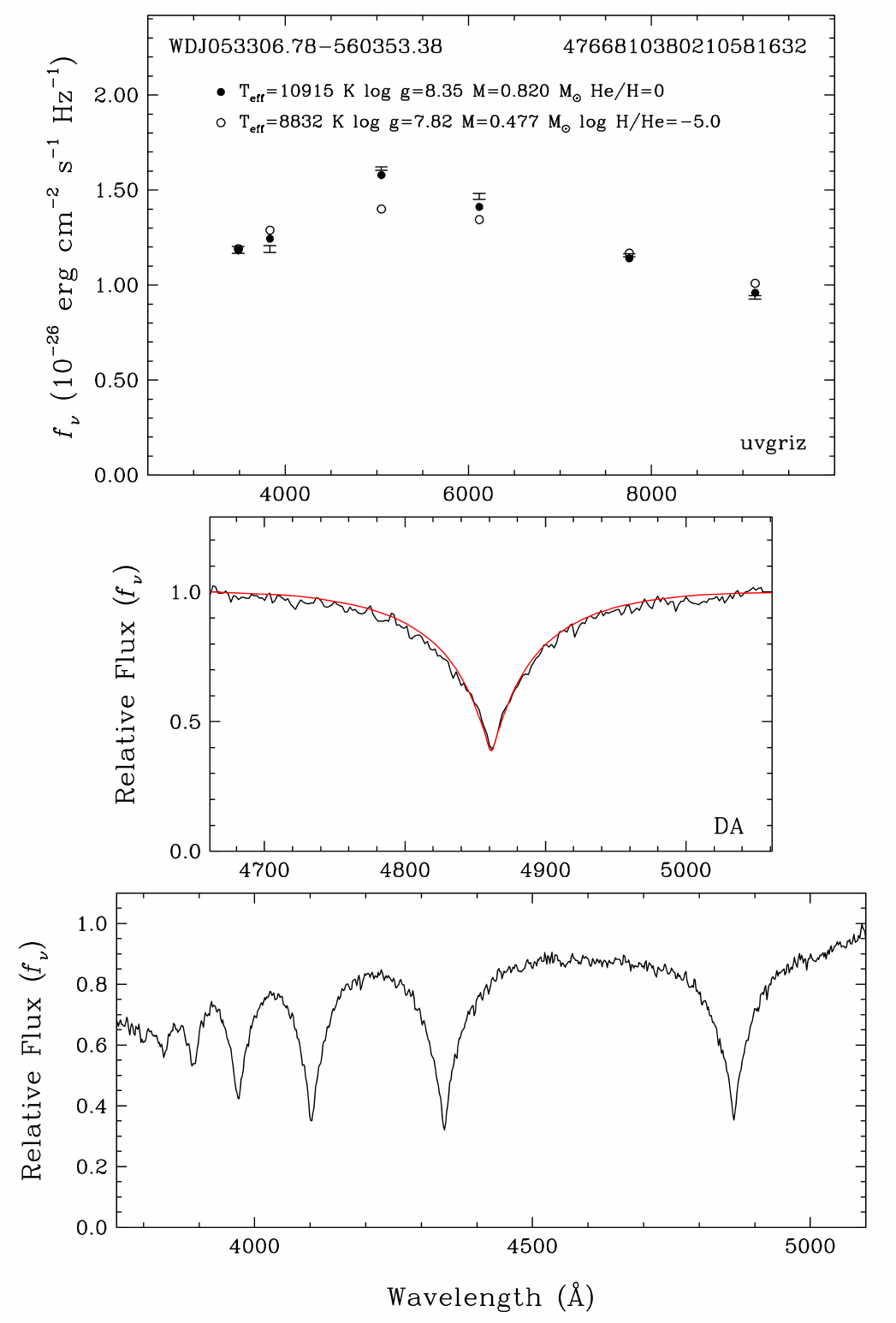} 
\caption{Model fits to the DA white dwarf WDJ053306.78$-$560353.38 (HE 0532$-$5605), which is a previously known ZZ Ceti pulsator in the PLATO
field. The top panel shows the best-fitting pure H (filled dots) and pure He (open circles) atmosphere white dwarf models to the Skymapper
photometry (error bars). This panel also includes the name and the \textit{Gaia} Source ID,  and the photometry used in the fitting.
The middle panel shows the predicted spectrum (red line) based on the pure-H solution, over-plotted on the observed spectrum (black line). The bottom panel presents the spectrum over a broader wavelength range.
}
\label{figda}
\end{figure}

As an example, Figure \ref{figda} shows our model fits to one of the pulsating ZZ Ceti stars known in the PLATO field, WDJ053306.78$-$560353.38 (HE 0532$-$5605).
The top panel shows the Skymapper $uvgriz$ photometry (error bars) along with the predicted fluxes from the best-fitting pure H (filled dots) and
pure He (open circles) atmosphere models. The labels in the same panel give the name, \textit{Gaia} source ID, and the photometry used in the fitting.
The photometry clearly favors the pure H solution, where a Balmer decrement is visible between the $uv$ and $g$ bands. The middle panel shows the
predicted spectrum based on the pure hydrogen solution (red line), along with the observed H$\alpha$ line. This is not a fit to the line profile,
we simply over-plot the predicted hydrogen line from the photometric fit. The bottom panel shows a broader spectral range. The photometric fit provides an excellent match to the H$\alpha$ line for this white dwarf, confirming that this is a pure hydrogen atmosphere white dwarf with $T_{\rm eff} = 10\,915 \pm 121$ K and $M=0.820 \pm 0.012~M_{\odot}$.

Tables~\ref{tab:DA} and \ref{tab:DB} in the Appendix present the white dwarf parameters derived in this work. 
These tables compile the targets fitted with pure-H (He/H = 0) and He-rich atmospheres with trace amounts of hydrogen (log H/He = –5). 
Each table lists \textit{Gaia} DR3 data (WD name, \textit{Gaia} source ID, RA, Dec, probability of being a white dwarf, and $G$ magnitude), PLATO Input Catalog (PIC) information (P magnitude and number of cameras), and the physical parameters ($T_{\rm eff}$, $\log g$, and mass) derived from our photometric fits. 
There are 118 DA stars in total, listed in Table~\ref{tab:DA}, with a subset (23) falling within the ZZ~Ceti instability strip; these objects will be prioritized for future follow-up studies (see Section \ref{DA_IS}). 
In addition, 6 stars classified as DB, DBA, DBAZ and 35 stars are DBV candidates (see Section \ref{DB_IS}) are included, with mixed atmosphere model solutions provided in Table~\ref{tab:DB}. 
The broader sample includes stars of other spectral types (e.g., DZ, DC, very hot WDs), but because these preliminary fits are not sufficiently robust — and in some cases may involve misclassifications such as hot subdwarfs — they are not published here. These preliminary estimates will instead remain available for internal use (e.g., target selection and follow-up planning) but are deliberately excluded from the paper to avoid confusion and ensure reliability of the published parameters.

While the Gaia DR3 XP spectra and the synthetic photometry from the \cite{2024A&A...682A...5V} catalog offer an important complementary resource for WD classification and parameter estimation, our analysis is based on an independent photometric fitting approach. Although the machine-learning classifications from XP photometry are generally reliable for DA white dwarfs, we noted inconsistencies for other spectral types, particularly for DB, DQ, and magnetic white dwarfs based on comparisons with follow-up spectroscopy. Since the method presented in \cite{2024A&A...682A...5V} also relies on photometric fitting techniques similar to ours, and adopts pure-He models for DBs, we opted for a more conservative approach. Our analysis includes both pure-H and mixed H/He models, which are more appropriate for DB white dwarfs, especially given that many ($\sim75$\%) DBs in the solar neighborhood also show trace amounts of H \citep[DBA spectral type,][]{2015A&A...583A..86K,kilic25}. This distinction is crucial, as our mixed-atmosphere models yield cooler temperatures for several DBV candidates, affecting their classification within the instability strip.

\section{Simulations}\label{sec:simulations}

As discussed in Sect.~\ref{sec:sample}, light curves of WD pulsators were simulated as part of the MOCKA catalogue \citep{2025A&A...694A.185J}.  While the analysis of MOCKA's simulated dataset focused on main sequence $g$-mode pulsators, the simulations showed principally that PLATO can detect pulsations with amplitudes $<0.1$ ppt and resolve closely spaced modes even at a 600s cadence while using input templates based on \textit{Kepler}, TESS, and ground-based light curves. In this section, we expand the analysis of MOCKA for WD pulsators. 

\subsection{Simulation setup}\label{sec:simulations_setup}

We use the open-source PLATO camera simulator \citep[\texttt{PlatoSim,}][]{2024A&A...681A..18J} to assess the feasibility of using PLATO for WD asteroseismology. For a comprehensive explanation on how the simulations were conducted (and which input parameters were used), we refer the reader to \citep{2025A&A...694A.185J}. In summary, we utilise the \texttt{PlatoSim} toolkit called \texttt{Platonium} to simulate photometric light curves using the on-board photometry \citep[][implemented into \texttt{PlatoSim}]{marchiori2019flight} for each camera and each a 90-d mission quarter. By default, \texttt{Platonium} use the standard \texttt{PlatoSim} input YAML file configured with the 'as expected' mission requirements. Post-processing is first done per camera-quarter light curve (to remove trends and photometric outliers), for then to create a single final light curve by averaging equal timings. Lastly, a 50 seconds sampling was used (i.e. twice the normal data rate of the N-CAMs) to increase the photometric precision while still allowing to all possible WD pulsation modes to be securely detected below the Nyquist frequency of $\delta \nu = 0.5/\delta t = 10,000 \,\mu\text{Hz}$ (or $864 \,\text{d}^{-1}$). 

\subsection{Simulation strategy}\label{sec:simulations_results}

As part of the MOCKA catalogue, 30 pulsating WD stars from the PLATO-CS input catalogue (c.f. Fig.~\ref{fig:cmd_WD}) were simulated: 10 DAV with pulsation modes extracted from \textit{Kepler} observations, and 10 DAV, 5 DBV, and 5 GW Vir stars with pulsation modes extracted from TESS observations. These 30 targets were selected based on being the richest pulsators, i.e., with the largest number of pulsation periods detected. While the pulsation modes were used to construct the injected template light curves, each pulsator was further simulated for ten different PLATO-CS catalogue stars across the four N-CAM visibilities. 
In the initial work by \citet{2024A&A...681A..18J}, 10 pulsating white dwarf stars from the PLATO-CS input catalogue were simulated. As an extension to this effort, we additionally selected and simulated ten more stars — thus increasing the sample to 20 in total — to ensure that five stars are observed by 6, 12, 18, and 24 co-pointing N-CAMs.

\begin{figure}
\centering
\includegraphics[width=\columnwidth]{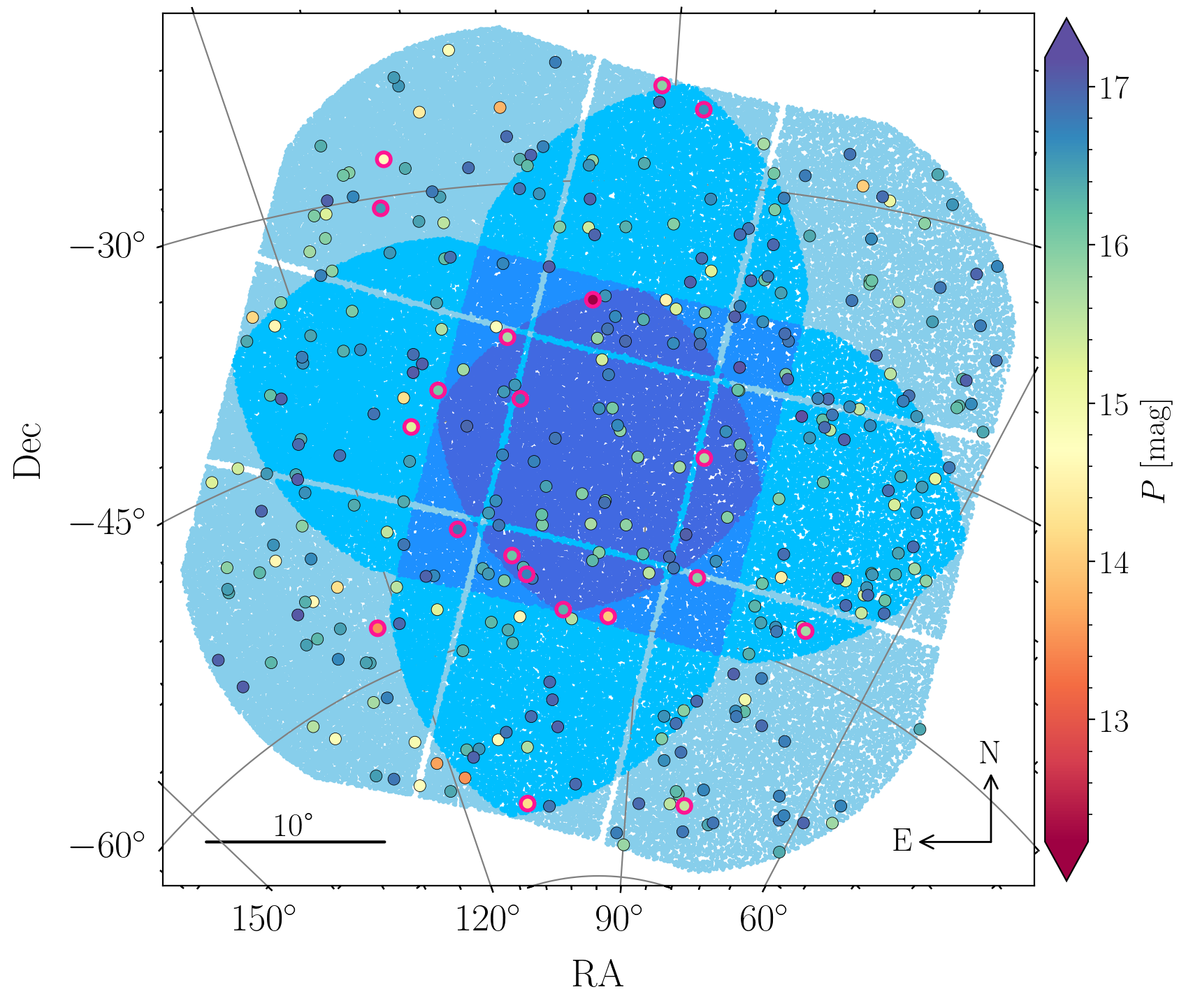}
\caption{The first PLATO pointing field in the South, LOPS2, shown in equatorial coordinates. The increasingly darker shade of blue illustrates the N-CAM overlap of $n_{\rm CAM} \in \{6, 12, 18, 24\}$, respectively. The color-coded circles (according to the PLATO N-CAM passband magnitude, $P$) show the WD candidates from Fig.~\ref{fig:cmd_WD}, while the pink circles show the 20 WD stars from the PLATO-CS catalogue used for simulations in this paper. This figure has been adapted from \cite{2025A&A...694A.185J}.}
\label{fig:LOPS2_WD}
\end{figure}

Thus, besides the N-CAM visibility, the $30\times20$ stars simulated were selected to cover a wide parameter space in $P$-band magnitude (from $\sim 12$ to $\sim 17$ mag), in random and systematic noise (unique for each of the 20 stars' location in the focal plane array), and levels of stellar contamination from nearby stars. Thus, all in all, realistically representing future observations. Figure~\ref{fig:LOPS2_WD} shows the LOPS2 overplotted with the WD candidates from the PLATO-CS input catalogue (color coded after PLATO magnitude). The 20 WD candidates used to generate all simulated light curves of this paper are illustrated with pink circles.

\subsection{Results for WD pulsators}\label{sec:simulations_results}

\begin{figure*}[t!]
\center
\includegraphics[width=1\columnwidth]{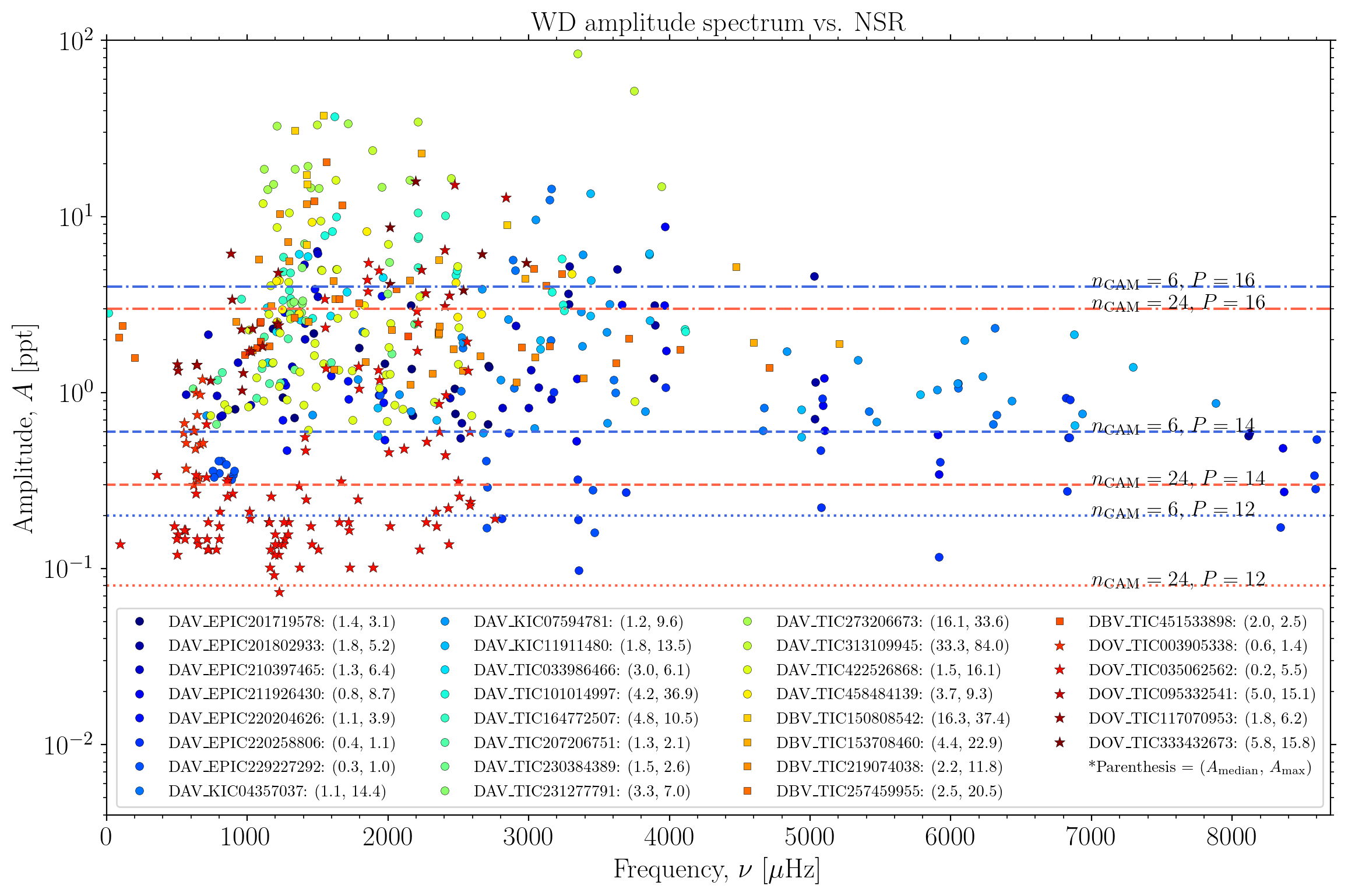}
\caption[]
{Pulsation amplitude vs. NSR diagram. Each previously observed pulsator has a unique colour-coding, while DAV, DBV, and GW Vir stars are marked with circles, squares, and stars. The horizontal lines represent general time domain noise limits for $n_{\rm CAM}$ of 6 (blue) and 24 (red) as a function of $G$-band magnitude corresponding to 12 (dotted), 14 (dashed), and 16 (dashed-dotted). Within the rounded brackets of the legend, we indicate respectively the median and maximum pulsation amplitudes for each target in units of ppt.} 
\label{fig:NSR}
\end{figure*}

To gain some basic insight into what kind of data quality PLATO will provide for pulsating WDs, we first investigate the underlying noise budget in the time domain. For our 30 mock stars, we investigate the noise-to-signal ratio (NSR) as a function of $P$-band magnitude and co-pointing cameras. We illustrate this in Fig.~\ref{fig:NSR} where the amplitude spectrum (shown with individual pulsation modes) of each WD candidate has a unique color-coding, while the pulsation modes of DAV, DBV, and GW Vir stars are marked with circles, squares, and stars, respectively. To first order, a given pulsation mode can only be detected if the pulsation amplitude is greater (with a few orders of magnitude) than the NSR level. Since PLATO's footprint directly map the camera visibility (see Fig.~\ref{fig:LOPS2_WD}), we provide six NSR median estimates as a reference (horizontal lines) across two N-CAM visibilities (6 and 24) and three different magnitudes (12, 14, and 16) covering the (bright, mean, and faint end) magnitude distribution of our WD sample. These NSR estimates were derived from MOCKA light curves \citep[c.f.][Fig.~C.1]{2025A&A...694A.185J}. 

As an example, it is evident from Fig.~\ref{fig:NSR} that for $n_{\rm CAM}=24$ and $P\leq12$, almost all pulsation modes will be detectable (notably those of the DAV star EPIC\,229227292 and the GW Vir star TIC\,035062562 i.e., GW Vir itself). Meanwhile, for $n_{\rm CAM}=6$ and $P\geq16$, only a fraction of the modes are expected to be detected. While this indicates the mode occurrence rates for WD pulsators, a thorough frequency analysis is needed to assess the underlying potential that PLATO will provide WD research, which we discuss in the following.

Considering now our simulated PLATO light curves, we overall find a good agreement in the pulsation mode detectability derived from amplitude spectra as compared to the (above-mentioned) estimates in the time domain. In Appendix \ref{app:amplitude_spectra}, we present as a showcase the amplitude spectra for three representative pulsating WDs simulated with {\tt PlatoSim}: the DAV TIC422526868 (G29-38, Fig. \ref{fig:amplitude_spectrum_DAV_TIC422526868}), the DBV TIC219074038 (GD358, Fig. \ref{fig:amplitude_spectrum_DBV_TIC219074038}), and the GW Vir star TIC035062562 (PG1159-035, Fig. \ref{fig:amplitude_spectrum_DOV_TIC035062562}). Each figure displays the (fast) Lomb–Scargle amplitude spectra of the injected (orange) and simulated (black) light curves for 20 mock realisations, illustrating the fidelity of the injected pulsation modes across different noise and visibility conditions. In particular, we highlight four conditions being the value of $n_{\rm CAM}$, the stellar magnitude, the stellar pollution rate (SPR; i.e. how much contaminating light that leaks into the target aperture mask on average), and the mean gnomonic radial distance the target star is away from the optical axis of each camera $\vartheta_{\rm OA}$. While $n_{\rm CAM}$ and the stellar magnitude set the base noise level, the SPR and $\vartheta_{\rm OA}$ are important noise contributors. In particular, a high SPR will naturally suppress the mode amplitudes, which is evident from each panel of Fig.~\ref{fig:amplitude_spectrum_DAV_TIC422526868}, \ref{fig:amplitude_spectrum_DBV_TIC219074038}, and \ref{fig:amplitude_spectrum_DOV_TIC035062562} for those target stars that have a non-zero SPR. On the other hand, the higher $\vartheta_{\rm OA}$ is, the lower the instrument transmission efficiency is and, thus, the lower the mode detectability is. 

Aside from the multiple noise sources that contribute to PLATO's overall noise budget, our analysis of all 30 pulsating stars simulated demonstrates that PLATO will be able to detect almost all pulsation modes in the bright end, and more than 50\% of the modes for all simulations, irrespective of the N-CAM visibility. The only exception is for heavily contaminated stars with $\text{SPR} > 10\%$ (e.g. see subpanel with benchmark star ID 17 in Fig.~.~\ref{fig:amplitude_spectrum_DAV_TIC422526868}, \ref{fig:amplitude_spectrum_DBV_TIC219074038}, and \ref{fig:amplitude_spectrum_DOV_TIC035062562}). Vanishing mode suppression is observed for $\text{SPR} < 6\%$ which agrees with the SPR threshold found by \cite{2025A&A...694A.185J}.

\begin{figure*}[t!]
\center
\includegraphics[width=0.9\columnwidth]{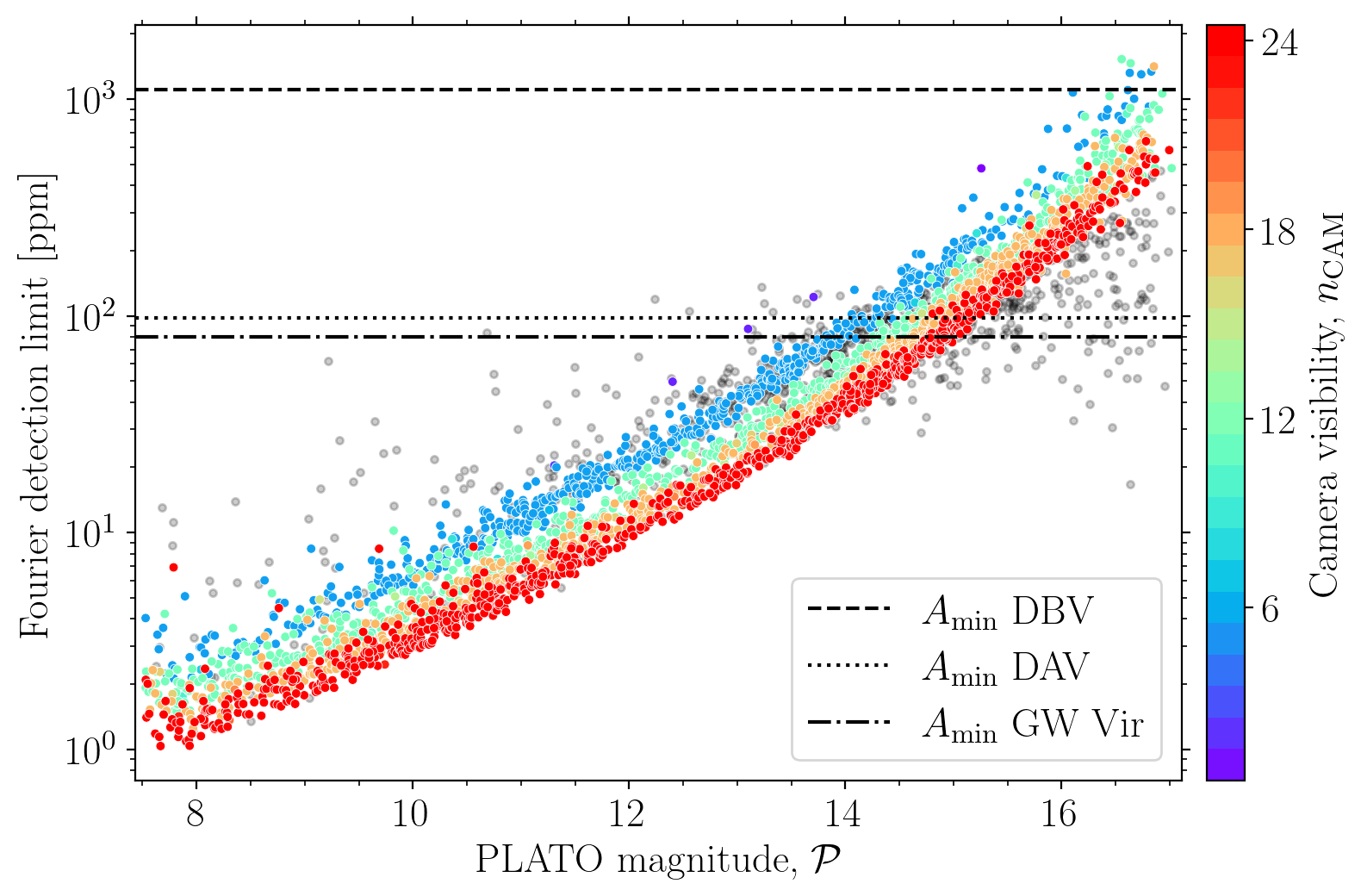}
\caption[]
{Fourier detection limit vs. PLATO magnitude diagram. This plot illustrates the general noise budget in the frequency domain enforced random and systematic noise sources. Each data point corresponds to the smallest amplitude extracted for each star (of a large sample of gravity mode pulsating stars) using prewhitening with a S/N stopping criterion. The horizontal lines show the minimum pulsation amplitude of the richest WD subclasses simulated: DBV (dashed), DAV (dotted), and GW Vir (dash-dotted). The black marker are star where $\text{SPR} > 6\%$ and hence subject to strong amplitude suppression. This figure has been adapted from \citep{2025A&A...694A.185J}.} 
\label{fig:fourier}
\end{figure*}

We quantify the above results in Fig.~\ref{fig:fourier}, which shows the noise budget in the frequency domain as a function of PLATO magnitude. Here, each data point corresponds to the smallest-amplitude mode extracted for a star using prewhitening with an S/N stopping criterion. Combined over a large sample of gravity mode pulsators, a clear noise structure is observed as a function of PLATO magnitude and with a strong dependency on N-CAM visibility. Given that the mode amplitudes vary from one WD subclass to another, we here show the minimum mode amplitude detected (horizontal lines) from each subclass of WD pulsator simulated, being DBV (dashed), DAV (dotted), and GW Vir (dash-dotted). 

As DBV stars generally exhibit the largest mode amplitudes among WD pulsators, we see that their typical minimum amplitude is detectable until $P<16$ even with 6 N-CAMs. On the other hand, DAV and GW Vir stars generally show smaller but similar mode amplitudes. With a typical minimum amplitude of $\sim 0.1\,\text{ppt}$ for both subclasses, with 6 N-CAMs, all mode amplitudes can be recovered up to a limiting magnitude of $P \lesssim 13.5$, while with 24 N-CAMs, all mode amplitudes can be recovered up to $P \lesssim 15$. It should be noted that these predictions depends on the SPR (clearly show with the black markers in Fig.~\ref{fig:fourier} that are stars with  $\text{SPR} > 6\%$) and  $\vartheta_{\rm OA}$ (which is the dominating cause for the spread of each noise budget curve representing $n_{\rm CAM} \in \{6, 12, 18, 24\}$). 

Figure~\ref{fig:fourier} furthermore illustrates the observational parameter space accessible for the different classes of pulsating WDs targeted by PLATO. Since the detectability limits depend directly on the intrinsic pulsation amplitudes of each subclass, the science return varies considerably between the different WD instability strips. DBV stars, which typically exhibit pulsation amplitudes in the range of $\sim0.05$--$0.3,\mathrm{ppt}$, remain detectable over the widest magnitude range and even under relatively poor N-CAM visibility conditions. This makes them particularly well suited for detailed asteroseismic analyses, mode stability studies, and investigations of plasmon neutrino cooling. Importantly, all DBV targets listed in Appendix~\ref{tableDBV} are expected to be recovered even for less optimal PLATO configurations.

GW Vir stars generally show lower amplitudes, typically between $\sim0.01$ and $0.15,\mathrm{ppt}$, and therefore require either brighter magnitudes or higher N-CAM visibility to recover their full pulsation spectra. Nevertheless, the simulations indicate that PLATO will recover the dominant pulsation content for the majority of known bright GW Vir stars, enabling detailed seismic studies of their internal chemical stratification, residual shell burning, and late-stage evolutionary processes connected to the PG1159 phase.

The DAV instability strip spans a particularly broad range of pulsation amplitudes and periods. Hot DAVs often exhibit low amplitudes of only $\sim0.001$--$0.015,\mathrm{ppt}$, whereas cooler ZZ Ceti stars can reach amplitudes up to $\sim0.3,\mathrm{ppt}$. Our simulations suggest that typical DAVs will be well recovered within the nominal PLATO magnitude range, especially for targets observed with moderate-to-high N-CAM visibility. This also includes ELMV and pre-ELMV pulsators, whose amplitudes typically range between $\sim0.001$ and $0.05,\mathrm{ppt}$. Consequently, all priority DAV-related targets listed in Appendix~\ref{tableDAV} are expected to remain detectable even under relatively unfavourable observing conditions. On the other hand, the situation becomes more challenging for massive and ultra-massive DAVs. Although these objects are of particular scientific interest due to their crystallized interiors and potential ONe-core compositions, their intrinsic faintness typically places them beyond $P \gtrsim 17$, where the PLATO Fourier detection limit becomes comparable to or larger than their expected pulsation amplitudes. As a result, recovering their complete mode spectra will likely only be feasible for the brightest and least contaminated systems. Similar limitations may also apply to low-amplitude DQV pulsators, whose amplitudes are expected to lie near the PLATO detection threshold for faint targets.

Overall, Fig.~\ref{fig:fourier} demonstrates that PLATO will provide excellent seismic capabilities for the majority of known bright pulsating WDs, while simultaneously defining the practical observational limits for the faintest and lowest-amplitude subclasses.

Based on the above-mentioned number statistics, the simulations presented in this work will serve as a solid benchmark for the selection of optimal WD targets for our first competitive GO proposal, while future \texttt{PlatoSim} simulations will continuously help shape each preparatory phase before each GO call.

\section{Conclusions and future prospects}


PLATO’s unprecedented capability to deliver two years of uninterrupted, high signal-to-noise light curves from its 26 cameras will revolutionize WD asteroseismology. By identifying common patterns in pulsation spectra of pulsating WDs, we will systematically compare observed frequencies with theoretical stellar models. This comparison will allow us to determine fundamental parameters such as WD mass through asymptotic period spacing analysis based on model grids spanning a broad range of masses and effective temperatures. In parallel, fitting individual pulsation periods with detailed models will provide deeper insights into internal structure, effective temperature, and stratification. These models will further yield seismological distances, which can be cross-validated against \textit{Gaia}’s precise astrometric measurements to test the reliability and robustness of asteroseismic models.
Moreover, the detection of rotational splitting in non-radial pulsation modes will enable us to constrain the internal rotation periods of WDs, probing angular momentum evolution. 
PLATO’s long-duration light curves will help uncover rare and poorly understood WD pulsator classes, such as ultra-massive WDs, ELMs, and hot DAVs, and refine our mapping of instability strips across the diverse WD spectral zoo (DAV, DBV, GW Vir).

We created a sample of 159 high-priority WD candidates ($G\,\leq\,17$) identified in PLATO’s Southern LOPS2 field using the PLATO-CS MOCKA catalogue and \textit{Gaia} DR3, with atmospheric parameters derived via photometric modeling. This includes 118 DAs (23 ZZ Ceti candidates), 41 non-DAs (35 DBV candidates).

Simulated observations using \texttt{PlatoSim} reveal that, for pulsation periods $G\leq12$ mag and with at least 24 co-pointing cameras ($n_{\rm CAM} = 24$), PLATO will detect nearly all relevant white dwarf pulsation modes. In contrast, with fewer cameras ($n_{\rm CAM} = 6$) and faint WDs ($G\geq17$ mag), only a fraction of the pulsation modes are expected to be detectable due to higher noise levels. These results underscore PLATO’s transformative potential to deliver a high-quality, space-based asteroseismic dataset for the white dwarf population, enabling unprecedented insights into the internal structure and evolution of stellar remnants.


Future spectroscopic and photometric observations of pulsating WDs are crucial to complement PLATO’s data and enable a more thorough characterization. Ongoing and forthcoming large-scale spectroscopic surveys, such as Sloan Digital Sky Survey V \citep[{\it SDSS-V};][]{2017arXiv171103234K} and 4MOST \citep{2019Msngr.175....3D}, will play a vital role in confirming spectroscopic classifications and expanding the pulsating WD sample size. 
In addition to the spectroscopic surveys, ongoing and upcoming photometric missions, including the TESS, alongside ground-based initiatives such as the Vera C. Rubin Observatory’s Legacy Survey of Space and Time \citep[{\it LSST};][]{2019ApJ...873..111I} and {\it BlackGEM} \citep[][]{2024PASP..136k5003G}, will provide complementary time-domain data. 
These combined efforts will improve our understanding of white dwarf variability, reveal new pulsation modes, and enable more precise stellar modeling. A coordinated survey of the LOPS2 region, supported by ground-based spectroscopy, will be essential to maximize the scientific return of PLATO’s white dwarf program. Together, these efforts will yield a rich, well-characterized sample of pulsating white dwarfs, offering new insights into the final stages of stellar evolution.

\backmatter

\bmhead{Supplementary information}

\bmhead{Acknowledgements}

This work presents results from the European Space Agency (ESA) space mission PLATO. The PLATO payload, the PLATO Ground Segment and PLATO data processing are joint developments of ESA and the PLATO mission consortium (PMC). Funding for the PMC is provided at national levels, in particular by countries participating in the PLATO Multilateral Agreement (Austria, Belgium, Czech Republic, Denmark, France, Germany, Italy, Netherlands, Portugal, Spain, Sweden, Switzerland, Norway, and United Kingdom) and institutions from Brazil. Members of the PLATO Consortium can be found at \url{https://platomission.com/}. The ESA PLATO mission website is \url{https://www.cosmos.esa.int/plato}. We thank the teams working for PLATO for all their work.
M.U. gratefully acknowledges funding from the Research Foundation Flanders (FWO) by means of a junior postdoctoral fellowship (grant agreement No. 1247624N). 
M.K. acknowledges support by the NSF under grant  AST-2205736, and the NASA under grants 80NSSC22K0479, 80NSSC24K0380, and 80NSSC24K0436.
P.S. acknowledges support from the Agencia Estatal de Investigaci\'on del Ministerio de Ciencia, Innovaci\'on y Universidades (MCIU/AEI) under grant ``Revolucionando el conocimiento de la evoluci\'on de estrellas poco masivas'' and the European Union NextGenerationEU/PRTR with reference CNS2023-143910 (DOI:10.13039/501100011033).
Zs.B. acknowledges the financial support of the KKP-137523 `SeismoLab' \'Elvonal grant of the Hungarian Research, Development and Innovation Office (NKFIH).
The research behind these results has received funding from the BELgian federal Science Policy Office (BELSPO) through PRODEX grant PLATO: ZKE2050-01-D01. We acknowledge the PLATO CS team, the simulation developers at KU Leuven, and ESA for making this mission possible.

\section*{Declarations}

\noindent\textbf{Conflict of interest/Competing interests}  
The authors declare that they have no conflict of interest. 

\noindent\textbf{Code availability}  
The codes used in this study are publicly available at \url{https://ivs-kuleuven.github.io/PlatoSim3/}.

\newpage

\begin{appendices}
\section{}

In this appendix, we present the identification of candidate pulsating white dwarfs within the LOPS2 sample based on their atmospheric parameters and location within the known instability strips of the different WD variability classes. Using the photometric solutions derived in this work, we compare the targets against the empirical instability-strip boundaries reported in the literature and assess the consistency of their spectral energy distributions with the expected atmospheric compositions. This analysis allows us to identify promising pulsating WD candidates for future photometric follow-up observations and for potential inclusion in PLATO Guest Observer programmes.

\section{Pulsating WD candidates in the LOPS2}

\subsection{DA instability strip}
\label{DA_IS}

\begin{figure}
\centering
\includegraphics[width=\columnwidth]{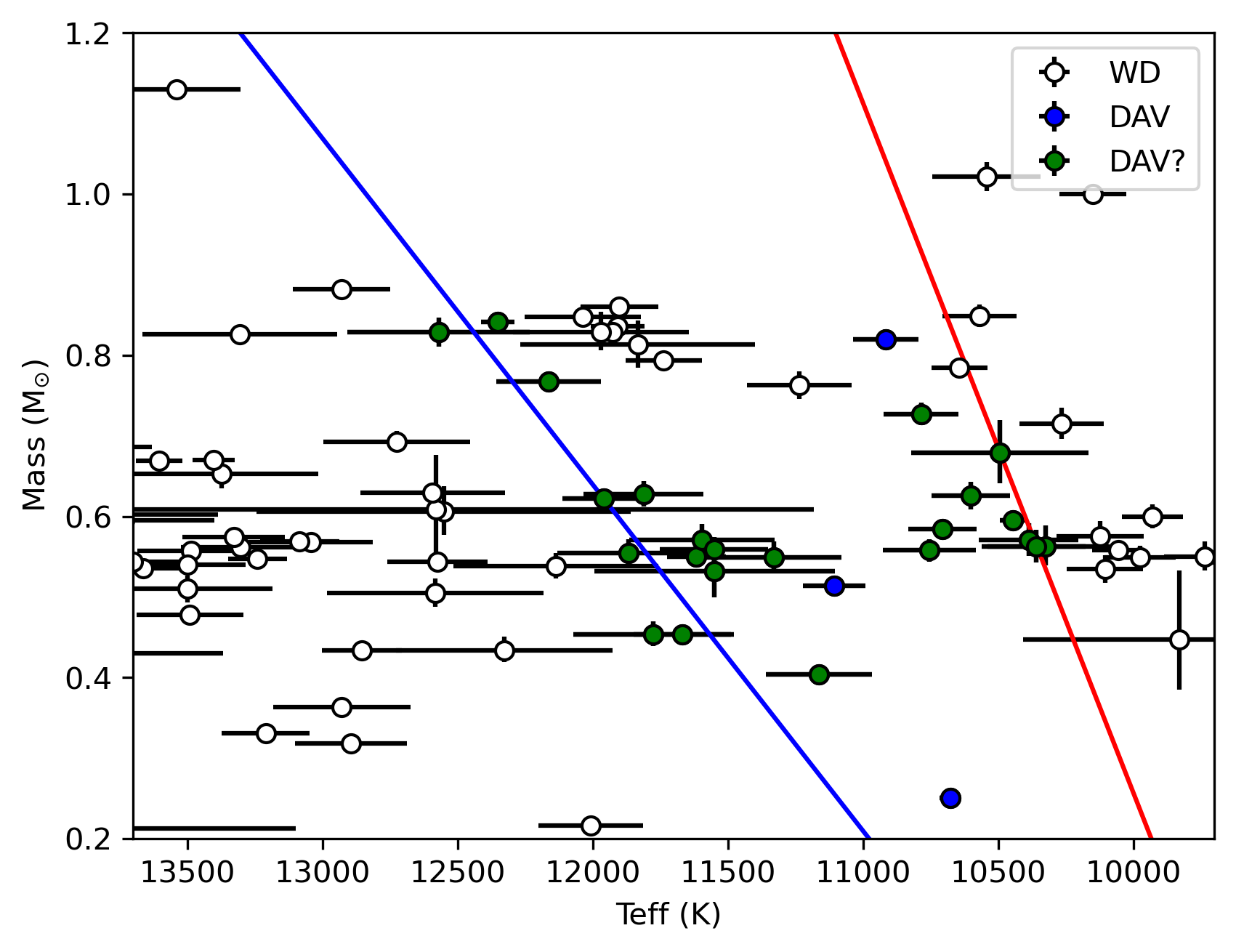}
\caption{Masses and effective temperatures for the PLATO WD sample assuming pure H atmospheres (open circles). 
The blue and red lines show the empirical boundaries of the ZZ Ceti instability strip from \citet{vincent20}.
Blue dots mark the three known ZZ Ceti stars in the sample, whereas the green dots mark the DAV candidates that fall
near or within the instability strip and also have photometric spectral energy distributions consistent with DA white dwarfs.}
\label{fig:DA_IS}
\end{figure}

Figure \ref{fig:DA_IS} shows the ZZ Ceti instability strip for PLATO WDs (open circles) under the assumption of pure H atmospheres. Here we exclude spectroscopically confirmed non-DAs and magnetic white dwarfs. The blue and red lines show the empirical boundaries of the ZZ Ceti instability strip from \citet{vincent20}. Blue dots mark the three previously known DAV pulsators, whereas the green dots mark the ZZ Ceti candidates that fall near or within the instability strip. We inspected the photometric spectral energy distributions of each candidate, and rejected the sources where the photometry clearly favors the He-atmosphere solution. At $T_{\rm eff}\sim11\,000$ K, DA WDs show a clear Balmer decrement, which is typically visible between the $uv$ and $g$ filters (see Figure \ref{figda}). We rejected 10 of the candidates as likely He-atmosphere objects; those are the open circles within the instability strip.

In total, we identify 23 ZZ Ceti candidates, where the spectral energy distributions are consistent with pure H atmospheres. Seven of these are spectroscopically confirmed to be DA WDs, including three pulsators. The known ZZ Ceti stars include the average mass WDJ045527.27$-$625844.61 with $T_{\rm eff}=11\,106 \pm 115$ K and $M=0.514 \pm 0.009~M_{\odot}$, the massive WDJ053306.78$-$560353.38 with $T_{\rm eff}=10\,915 \pm 121$ K and $M=0.820 \pm 0.012~M_{\odot}$, and the apparently low-mass WDJ071114.41$-$251815.05 with $T_{\rm eff}=10\,677 \pm 40$ K and $M=0.251 \pm 0.001~M_{\odot}$.

We also identify the DA white dwarf WDJ053203.91$-$653609.91 with $T_{\rm eff} = 11,330 \pm 250$ K and $M = 0.549 \pm0.02~M{\odot}$, and the candidate ZZ Ceti WDJ060052.91$-$463041.11, whose atmospheric parameters vary between two solutions: a low-temperature, low-mass fit with $T_{\rm eff} = 9857 \pm100$ K and $M = 0.261 \pm 0.004M_{\odot}$, and a higher-temperature solution with $T_{\rm eff} = 13,210 \pm 162$ K and $M = 0.331 \pm 0.006M_{\odot}$.
 
WDJ045527.27$-$625844.61 shows 4 pulsation modes in TESS with periods ranging from 126.8 to 320.8 s. Performing asteroseismology
based on these four modes, \citet{2022MNRAS.511.1574R} derive a mass of $0.542~M_\odot$, which is similar to the mass that we obtain from its
spectral energy distribution and \textit{Gaia} parallax. 
Similarly, WDJ053203.91$-$653609.91 exhibits pulsations with periods of 275.49 s and 411.28 s, and has an effective temperature of $12,390$ K and a mass of approximately $0.66 M_\odot$ \citep{2022MNRAS.511.1574R}. WDJ060052.91$-$463041.11 shows a pulsation period of 268.45 s but presents a degenerate solution in its atmospheric parameters, with one fit yielding $T_{\rm eff} \approx 12,150$ K and $M \approx 0.493 M_\odot$, and an alternative cooler solution at $T_{\rm eff} \approx 9,350$ K with $M \approx 0.303 M_\odot$ \citep{2022MNRAS.511.1574R}.
WDJ053306.78$-$560353.38 shows 8 modes ranging from 522 to 724 s \citep{castanheira09},
which is consistent with a pulsating massive DAV. On the other hand, based on a single detected mode at 278 s in TESS data, \citet{2022MNRAS.511.1574R}
derive a mass of $0.632~M_\odot$ for WDJ071114.41$-$251815.05, which is inconsistent with our mass estimate.
Note that \citet{2021MNRAS.508.3877G} also find a mass of $\approx0.3~M_\odot$ based on \textit{Gaia} photometry and astrometry for the same star.
WDJ071114.41$-$251815.05 is clearly over-luminous. The discrepancy between the asteroseismological and photometric mass estimates can be
resolved if this is a binary system, where the extra light from the companion results in an inflated radius estimate and a lower mass.
Note that the 278 s mode is significantly shorter than the typical pulsation modes seen in ELM WDs \citep{hermes12,kilic15}. Hence,
WDJ071114.41$-$251815.05 likely represents a pulsating DA with a WD companion. 
We provide the list of the remaining 23 ZZ Ceti candidates
in Table \ref{tableDAV}; 19 of these lack spectroscopy. We plan on obtaining follow-up spectroscopy to confirm their nature and their suitability
for PLATO observations.




\begin{table}
\scriptsize
\caption{DAV Candidates in the PLATO LOPS2 field.}   
\label{tableDAV}      
\centering                          
\begin{tabular}{c c c c}        
\hline\hline                 
Name & \textit{Gaia} Source ID & $T_{\rm eff}$ (K) & Mass ($M_\odot$) \\
\hline
WDJ041930.81$-$352046.73 & 4870090943381248128 & 10602 & 0.626 \\
WDJ045359.85$-$432247.75 & 4811871287295222016 & 11162 & 0.404 \\
WDJ045405.87$-$342505.71 & 4873109308958466432 & 11960 & 0.622 \\
WDJ045551.33$-$705541.81 & 4654695103690489856 & 10324 & 0.563 \\
WDJ052518.96$-$302018.13 & 2905372374461903488 & 11616 & 0.550 \\
WDJ052548.76$-$671611.58 & 4658691072554066560 & 10493 & 0.679 \\
WDJ052912.10$-$430334.49 & 4805782462481529600 & 11669 & 0.454 \\
WDJ053100.49$-$455801.14 & 4799224635833122304 & 10387 & 0.571 \\
WDJ053203.91$-$653609.91 & 4660769493122896128 & 11330 & 0.549 \\
WDJ053939.69$-$664157.85 & 4659528969150132096 & 11868 & 0.555 \\
WDJ054458.33$-$401940.13 & 4805139866655106944 & 11596 & 0.571 \\
WDJ055428.05$-$402555.11 & 4804209684114310784 & 12570 & 0.829 \\
WDJ060013.07$-$545516.04 & 5500308124837433984 & 10785 & 0.727 \\
WDJ064802.46$-$483039.16 & 5551402808137892864 & 11551 & 0.559 \\
WDJ071051.17$-$342107.14 & 5602672550010235008 & 10707 & 0.584 \\
WDJ071751.88$-$282928.84 & 5606764386835758720 & 10444 & 0.595 \\
WDJ072717.48$-$332838.16 & 5592485716514387584 & 11775 & 0.454 \\
WDJ073338.28$-$281623.17 & 5611744006283630720 & 12351 & 0.841 \\
WDJ074613.81$-$302550.93 & 5598661329740179712 & 10754 & 0.558 \\
WDJ075643.91$-$545348.89 & 5296195267296378112 & 11549 & 0.532 \\
WDJ083246.58$-$574125.49 & 5315128548292806656 & 11812 & 0.628 \\
WDJ083759.16$-$501745.76 & 5322090003089341440 & 12163 & 0.767 \\
WDJ084215.26$-$555628.51 & 5316824544982104960 & 10358 & 0.563 \\
\hline
\end{tabular}
\end{table}

\subsection{DB instability strip}
\label{DB_IS}

\begin{figure}
\centering
\includegraphics[width=\columnwidth]{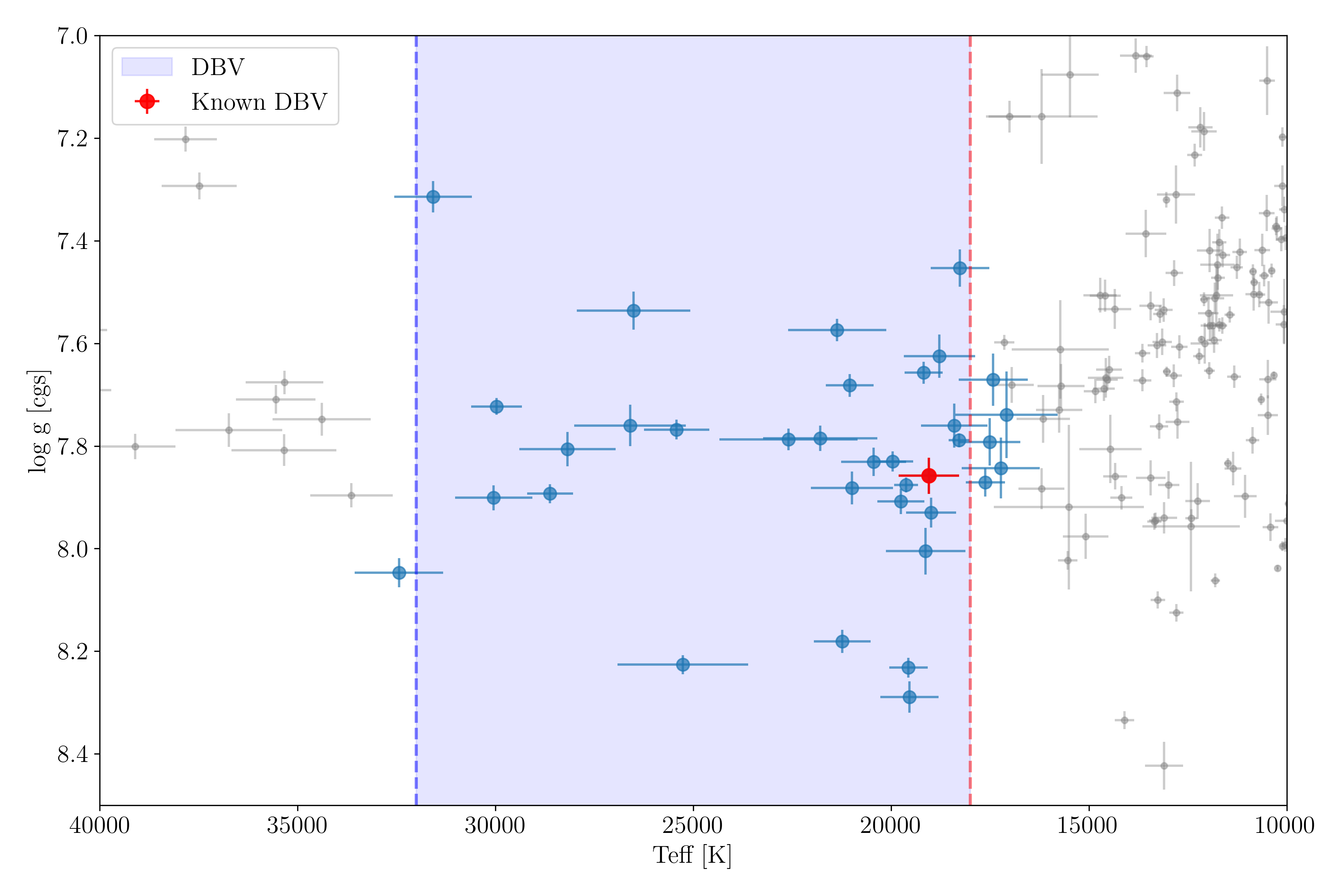}
\caption{Location of the PLATO white dwarf sample in the effective temperature–surface gravity plane, modeled under the assumption of He-rich atmospheres. The blue and red dashed lines indicate the extended empirical boundaries of the DB instability strip based on \citet{2019A&ARv..27....7C,2022ApJ...927..158V}. Blue points highlight stars located within or close to the strip. The known DBV WDJ052330.80−472219.55 is marked with a red dot.}
\label{fig:DB_IS}
\end{figure}

Figure \ref{fig:DB_IS} displays the location of PLATO DB white dwarfs modeled under the assumption of He-rich atmospheres (with traces of hydrogen)
in the $T_{\rm eff}$–surface gravity plane. As for the DA analysis, we removed spectroscopically confirmed magnetic WDs.
The blue and red dashed lines trace the empirical boundaries of the DB instability strip, which we extend using the work of \citet{2019A&ARv..27....7C} and \citet{2022ApJ...927..158V}. Blue points mark stars that fall within or near the instability strip. The only currently known
DBV in the PLATO field, WDJ052330.80$−$472219.55, is highlighted with a red dot; its location within the strip provides an important anchor point
for our analysis. 
In total, we identify 35 DBV candidates, which are listed in Table~\ref{tableDBV}.
These stars have effective temperatures ranging from $\sim17\,000$ K to 32\,500 K and span a broad mass range of about $0.40-0.71 M_\odot$. Unfortunately, the photometric spectral energy distributions do not allow us to distinguish
between DA and DB white dwarfs at these temperatures. Hence, follow-up spectroscopy is required to confirm DB white dwarfs in this sample.

The only confirmed DBV in our sample, WDJ052330.80−472219.55 (EC~05221$-$4725), was identified as a pulsating DB white dwarf by \citet{2009MNRAS.397..453K}, who reported clear variability with a dominant period near 15 minutes. Their time-series photometry revealed a persistent frequency around $\sim$890~s and additional peaks near $\sim$810~s, alongside evidence for amplitude modulation and beating effects, suggesting a complex pulsation spectrum worthy of further study. Upcoming long-baseline, high-precision light curves from PLATO will be extremely valuable for resolving its rich frequency spectrum and assessing mode stability.

\begin{table}
\scriptsize
\caption{DBV Candidates in the PLATO LOPS2 field.}   
\label{tableDBV}      
\centering                          
\begin{tabular}{c c c c}        
\hline\hline                 
Name & \textit{Gaia} source ID & $T_{\rm eff}$ (K) & Mass ($M_\odot$) \\
\hline

WDJ041724.58-673236.77 & 4668242152077665024 & 17419 & 0.428 \\ 
WDJ042610.85-465847.47 & 4789144175789941888 & 20442 & 0.51 \\
WDJ043151.06-532929.85 & 4780712394777586560 & 28181 & 0.514 \\
WDJ050510.82-613924.40 & 4760710942535282560 & 21235 & 0.709 \\
WDJ053423.94-381444.86 & 4808986478021189120 & 26509 & 0.401 \\
WDJ054007.06-651807.95 & 4660085867450146560 & 17084 & 0.458 \\
WDJ054657.41-305153.89 & 2903068214703934848 & 25266 & 0.742 \\
WDJ055635.50-561006.57 & 4767225652008399104 & 19561 & 0.739 \\
WDJ060612.38-723414.20 & 5266145236552677632 & 18259 & 0.348 \\
WDJ060740.96-285302.51 & 2897810350100517504 & 18283 & 0.484 \\
WDJ061109.12-703921.89 & 5278615205944625664 & 26597 & 0.49 \\
WDJ061238.45-325357.56 & 2892795821161588736 & 19180 & 0.427 \\
WDJ061524.39-384823.63 & 5574654180530308736 & 25422 & 0.491 \\
WDJ061856.82-372607.54 & 5575200775248718976 & 22594 & 0.493 \\
WDJ062000.90-233500.19 & 2936643760765256576 & 21047 & 0.442 \\
WDJ063814.10-694335.39 & 5278903110486861312 & 32439 & 0.646 \\
WDJ063845.14-470141.00 & 5552665081846145024 & 17512 & 0.484 \\
WDJ064648.90-525349.76 & 5498647381243346176 & 18988 & 0.558 \\
WDJ064747.51-270813.49 & 2919284671187233280 & 28620 & 0.557 \\
WDJ065112.06-440937.57 & 5562002134589548032 & 18778 & 0.413 \\
WDJ065703.65-433414.37 & 5559127976835354624 & 18403 & 0.471 \\
WDJ065957.09-291701.35 & 5608058473357824128 & 21363 & 0.4 \\
WDJ070918.12-302728.87 & 5604654870693369984 & 29970 & 0.481 \\
WDJ073920.28-262915.39 & 5613838438495596032 & 19624 & 0.53 \\
WDJ074250.12-544920.12 & 5488182813687216000 & 19127 & 0.6 \\
WDJ074521.91-432803.38 & 5532479697634251776 & 19536 & 0.775 \\
WDJ081612.57-402513.53 & 5539643909237818880 & 21792 & 0.49 \\
WDJ081827.11-460309.83 & 5519761028164909568 & 17228 & 0.508 \\
WDJ082109.21-500114.40 & 5514870018179554560 & 31577 & 0.343 \\
WDJ083341.91-655530.74 & 5272852979034659072 & 19752 & 0.547 \\
WDJ083942.55-535949.14 & 5317871245692348032 & 20988 & 0.536 \\
WDJ085424.48-491724.58 & 5328333889104483712 & 30045 & 0.564 \\
WDJ041909.91-535045.31 & 4779379172505736832 & 17617 & 0.524 \\
WDJ042211.36-474142.02 & 4788896304637763072 & 19960 & 0.508 \\
WDJ052330.80-472219.55 & 4797409582653745152 & 19045 & 0.52 \\

\hline
\end{tabular}
\end{table}

\section{Other Types of Variable White Dwarfs in the LOPS2}

Beyond pulsating white dwarfs, we also investigated other types of variable white dwarfs in our sample. Using the \texttt{TESS-Localizer} tool \citep{2023AJ....165..141H}, we extracted and classified light curves from TESS-observed targets, identifying 24 variable white dwarfs.
These variables exhibit a wide range of behaviors, including short-period binary systems, eclipsing binaries, spotted white dwarfs, candidates for ELMVs, and potential cataclysmic variables. The variability observed is often attributed to rotational modulation or binary interactions, with periods ranging from less than one hour to several tens of hours.
Approximately 15 of these variable white dwarfs coincide with targets studied in recent works \citep[e.g.,][]{2024ApJ...974..314O,2024MNRAS.528.6056H}, which focus on magnetic white dwarfs and other notable variable types.
Within this sample, we confirmed three cataclysmic variables, namely TIC\,364266095 \citep[Var Ret2005,][]{2023AJ....166..131T}, TIC\,172467837 \citep[PU CMa,][]{2003PASP..115...37T,2003MNRAS.339..861K}, and TIC\,25133286 \citep[VW Hyi,][]{1974MNRAS.166..673W}. 
We identified two short-period binaries hosting extremely low-mass white dwarfs: TIC\,278861557 (P$_{\rm orb}$ = 3.12,hr; J0642$-$5605) \citep{2020ApJ...894...53K} and TIC\,310478036 (P$_{\rm orb}$ = 2.32,hr; SDSS,J083417.21$-$652423.2) \citep{2025A&A...694A.246A}. In addition, TIC\,279321596 exhibits spot modulation, while TIC\,430591603 is a candidate ELMV white dwarf warranting further follow-up observations.

Finally, thanks to its blue bandpass and fast cadence, PLATO could uncover new transiting systems among white dwarfs if monitored. Using Gaia EDR3 data \citep{2021MNRAS.508.3877G}, \citet{2024MNRAS.535.1778E} identified two promising targets in the LOPS2 field: RXJ0623.2$-$3741 (TIC\,237397195), a metal-polluted white dwarf and strong planet-host candidate \citep{2019MNRAS.487.3470P}, and $\epsilon$\,Ret\,B (TIC\,684987240), the white dwarf companion to a known planet-hosting K subgiant \citep{2001ApJ...555..410B,2007A&A...469..755M}. The authors emphasize that white dwarfs in LOPS2 can achieve photometric precision comparable to PIC stars, and that deeper transits around white dwarfs make even fainter targets suitable for exoplanet detection.

\section{LOPS2 target list}\label{app:tables}

\onecolumn
\begin{landscape}

\small\tabcolsep=2pt

\centering

\begin{longtable}{ccccccccccccc}

\caption{
Parameters of DA white dwarfs modeled with pure-hydrogen atmospheres are presented, incorporating Gaia DR3 data, PLATO Input Catalog information, and photometric fit results, with targets ordered by increasing right ascension.
} \\

\toprule
Name & Gaia ID & RA [deg] & Dec [deg] & Pwd & $G$ [mag] & $P$ [mag] & $n_{\rm CAM}$ & Abundance & S. Type &  $T_{\rm eff}$ [K] & $\log g$ [cm\,s$^{-2}$] & $M$ [$M_\odot$] \\
\midrule
\endfirsthead

\toprule
Name & Gaia ID & RA [deg] & Dec [deg] & Pwd & $G$ [mag] & $P$ [mag] & $n_{\rm CAM}$ & Abundance & S. Type &  $T_{\rm eff}$ [K] & $\log g$ [cm\,s$^{-2}$] & $M$ [$M_\odot$] \\
\midrule
\endhead
\midrule
\multicolumn{13}{r}{{Continued on next page}} \\
\midrule
\endfoot

\bottomrule
\endlastfoot
WDJ033434.32-640057.11 & 4673757096243879168 & 53.64349 & -64.01608 & 0.03 & 14.03 & 13.88 & 6 & [He/H=0] & DA & $26552^{+707}_{-707}$ & $7.250^{+0.029}_{-0.027}$ & $0.343^{+0.011}_{-0.011}$ \\
WDJ033823.93-575826.83 & 4729626958543290368 & 54.59983 & -57.97409 & 0.01 & 16.33 & 16.47 & 6 & [He/H=0] & DA & $31370^{+956}_{-956}$ & $5.272^{+0.125}_{-0.100}$ & $0.235^{+0.046}_{-0.009}$ \\
WDJ034222.37-670935.88 & 4667932055438729728 & 55.59395 & -67.15954 & 1.00 & 15.72 & 15.78 & 6 & [He/H=0] & DA & $15219^{+272}_{-272}$ & $8.258^{+0.015}_{-0.014}$ & $0.771^{+0.013}_{-0.012}$ \\
WDJ035938.19-511839.83 & 4828757999190736128 & 59.90907 & -51.31099 & 1.00 & 15.76 & 15.89 & 12 & [He/H=0] & DA & $22359^{+300}_{-299}$ & $7.827^{+0.013}_{-0.012}$ & $0.539^{+0.008}_{-0.009}$ \\
WDJ040530.09-412109.11 & 4842194473663089920 & 61.37526 & -41.35250 & 1.00 & 16.29 & 16.41 & 6 & [He/H=0] & DA & $21502^{+452}_{-451}$ & $7.856^{+0.018}_{-0.018}$ & $0.551^{+0.013}_{-0.013}$ \\
WDJ040532.88-505557.89 & 4828681514413122944 & 61.38920 & -50.93269 & 0.99 & 15.81 & 15.76 & 12 & [He/H=0] & DA & $9607^{+123}_{-123}$ & $7.902^{+0.020}_{-0.020}$ & $0.543^{+0.015}_{-0.015}$ \\
WDJ040935.24-494219.28 & 4830333221396620032 & 62.39712 & -49.70509 & 1.00 & 16.75 & 16.92 & 12 & [He/H=0] & DA & $31531^{+938}_{-938}$ & $7.833^{+0.035}_{-0.033}$ & $0.565^{+0.023}_{-0.021}$ \\
WDJ041013.13-441053.05 & 4838637798361005696 & 62.55465 & -44.18141 & 1.00 & 16.64 & 16.82 & 12 & [He/H=0] & DA & $32618^{+593}_{-593}$ & $7.705^{+0.023}_{-0.022}$ & $0.511^{+0.013}_{-0.012}$ \\
WDJ041110.32-514651.04 & 4780544792270137088 & 62.79285 & -51.78109 & 1.00 & 16.18 & 16.32 & 12 & [He/H=0] & DA & $25151^{+419}_{-419}$ & $7.749^{+0.019}_{-0.018}$ & $0.510^{+0.012}_{-0.011}$ \\
WDJ041114.18-370053.94 & 4845518095450995328 & 62.80898 & -37.01495 & 1.00 & 16.75 & 16.87 & 6 & [He/H=0] & DA & $20796^{+419}_{-420}$ & $7.977^{+0.018}_{-0.019}$ & $0.613^{+0.014}_{-0.014}$ \\
WDJ041121.16-322614.82 & 4882864622796659072 & 62.83788 & -32.43787 & 0.06 & 16.13 & 15.92 & 6 & [He/H=0] & DAM & $15182^{+201}_{-201}$ & $7.602^{+0.022}_{-0.019}$ & $0.416^{+0.012}_{-0.011}$ \\
WDJ041250.75-451012.52 & 4837423353408638080 & 63.21214 & -45.16961 & 1.00 & 15.14 & 15.18 & 12 & [He/H=0] & DA & $13084^{+146}_{-147}$ & $7.931^{+0.008}_{-0.008}$ & $0.569^{+0.007}_{-0.007}$ \\
WDJ041337.30-502808.71 & 4782203646078747648 & 63.40546 & -50.46942 & 1.00 & 16.37 & 16.47 & 12 & [He/H=0] & DA & $19330^{+286}_{-285}$ & $7.829^{+0.015}_{-0.015}$ & $0.531^{+0.011}_{-0.010}$ \\
WDJ041539.80-402232.69 & 4841120770494891520 & 63.91592 & -40.37567 & 1.00 & 16.09 & 16.28 & 6 & [He/H=0] & DA & $34725^{+2317}_{-2316}$ & $7.428^{+0.065}_{-0.059}$ & $0.421^{+0.028}_{-0.026}$ \\
WDJ041602.97-403212.44 & 4841115341656167552 & 64.01268 & -40.53702 & 1.00 & 16.11 & 16.22 & 6 & [He/H=0] & DA & $20144^{+408}_{-408}$ & $7.945^{+0.018}_{-0.017}$ & $0.594^{+0.014}_{-0.013}$ \\
WDJ041630.04-591757.19 & 4678664766393827328 & 64.12478 & -59.30000 & 1.00 & 12.48 & 12.53 & 6 & [He/H=0] & DA & $13490^{+199}_{-198}$ & $7.754^{+0.014}_{-0.013}$ & $0.478^{+0.010}_{-0.009}$ \\
WDJ041711.36-545747.22 & 4778802753534553472 & 64.29762 & -54.96300 & 0.99 & 15.32 & 15.48 & 6 & [He/H=0] & DA & $28219^{+463}_{-463}$ & $6.943^{+0.018}_{-0.017}$ & $0.266^{+0.006}_{-0.005}$ \\
WDJ041804.15-384519.53 & 4844370755067765504 & 64.51754 & -38.75520 & 1.00 & 16.15 & 16.24 & 6 & [He/H=0] & DA & $18788^{+267}_{-266}$ & $7.902^{+0.014}_{-0.013}$ & $0.567^{+0.010}_{-0.009}$ \\
WDJ041921.96-302643.10 & 4884691594508544128 & 64.84196 & -30.44520 & 1.00 & 16.58 & 16.67 & 6 & [He/H=0] & DA & $18122^{+250}_{-250}$ & $7.859^{+0.014}_{-0.013}$ & $0.543^{+0.010}_{-0.009}$ \\
WDJ041924.71-531917.10 & 4779427928974390272 & 64.85283 & -53.32129 & 1.00 & 16.42 & 16.57 & 12 & [He/H=0] & DA & $25041^{+416}_{-415}$ & $7.728^{+0.019}_{-0.018}$ & $0.500^{+0.012}_{-0.011}$ \\
WDJ041930.81-352046.73 & 4870090943381248128 & 64.87816 & -35.34649 & 1.00 & 16.80 & 16.78 & 6 & [He/H=0] & DA & $10602^{+146}_{-145}$ & $8.041^{+0.020}_{-0.020}$ & $0.626^{+0.017}_{-0.017}$ \\
WDJ041955.84-522122.89 & 4781732470986533120 & 64.98248 & -52.35632 & 1.00 & 16.13 & 16.20 & 12 & [He/H=0] & DA & $15612^{+193}_{-192}$ & $7.861^{+0.012}_{-0.012}$ & $0.538^{+0.008}_{-0.009}$ \\
WDJ042357.67-455042.27 & 4789516154317811456 & 65.98968 & -45.84736 & 0.99 & 16.67 & 16.41 & 12 & [He/H=0] & DA & $5562^{+32}_{-32}$ & $7.974^{+0.013}_{-0.013}$ & $0.570^{+0.010}_{-0.011}$ \\
WDJ043133.39-404735.67 & 4863952954079430272 & 67.88915 & -40.79323 & 1.00 & 16.71 & 16.89 & 12 & [He/H=0] & DA & $30476^{+557}_{-557}$ & $7.764^{+0.023}_{-0.023}$ & $0.531^{+0.014}_{-0.014}$ \\
WDJ043254.41-325314.36 & 4871521270570577152 & 68.22689 & -32.88711 & 0.03 & 16.50 & 16.34 & 6 & [He/H=0] & DA & $7029^{+87}_{-87}$ & $5.701^{+0.024}_{-0.029}$ & $0.160^{+0.004}_{-0.007}$ \\
WDJ043255.87-355729.04 & 4867574023826934272 & 68.23443 & -35.95788 & 1.00 & 16.80 & 16.49 & 6 & [He/H=0] & DA & $4919^{+31}_{-30}$ & $7.843^{+0.016}_{-0.017}$ & $0.488^{+0.013}_{-0.012}$ \\
WDJ043333.60-275324.86 & 4891567154251463680 & 68.39196 & -27.89027 & 1.00 & 16.59 & 16.30 & 6 & [He/H=0] & DA & $5083^{+15}_{-14}$ & $8.004^{+0.006}_{-0.006}$ & $0.583^{+0.005}_{-0.005}$ \\
WDJ043704.01-572821.62 & 4775127807717312640 & 69.26679 & -57.47228 & 1.00 & 15.63 & 15.74 & 12 & [He/H=0] & DA & $20100^{+448}_{-448}$ & $8.075^{+0.018}_{-0.018}$ & $0.667^{+0.015}_{-0.014}$ \\
WDJ043859.53-285614.80 & 4879382984867446144 & 69.74802 & -28.93757 & 1.00 & 16.48 & 16.58 & 6 & [He/H=0] & DA & $19353^{+231}_{-232}$ & $7.857^{+0.012}_{-0.011}$ & $0.546^{+0.008}_{-0.008}$ \\
WDJ043915.39-571900.57 & 4774755245074357120 & 69.81406 & -57.31677 & 0.99 & 15.63 & 15.79 & 12 & [He/H=0] & DA & $29018^{+754}_{-754}$ & $7.119^{+0.029}_{-0.027}$ & $0.312^{+0.012}_{-0.011}$ \\
WDJ044137.51-485503.02 & 4785027741695161728 & 70.40635 & -48.91746 & 0.01 & 15.22 & 15.40 & 12 & [He/H=0] & DA & $38956^{+1581}_{-1582}$ & $5.274^{+0.053}_{-nan}$ & $0.200^{+0.004}_{-0.464}$ \\
WDJ044550.77-385538.25 & 4818044770206932864 & 71.46149 & -38.92727 & 1.00 & 16.76 & 16.96 & 12 & [He/H=0] & DA & $44971^{+5476}_{-5476}$ & $7.722^{+0.069}_{-0.061}$ & $0.550^{+0.042}_{-0.035}$ \\
WDJ044901.96-423414.72 & 4814969195665821568 & 72.25826 & -42.57066 & 1.00 & 16.58 & 16.65 & 12 & [He/H=0] & DA & $15937^{+332}_{-331}$ & $7.922^{+0.019}_{-0.018}$ & $0.571^{+0.015}_{-0.014}$ \\
WDJ044957.96-474959.45 & 4786446146054306048 & 72.49150 & -47.83327 & 1.00 & 16.04 & 16.14 & 12 & [He/H=0] & DA & $18533^{+225}_{-224}$ & $7.837^{+0.012}_{-0.012}$ & $0.533^{+0.009}_{-0.008}$ \\
WDJ045151.11-730250.29 & 4652838475220068992 & 72.96524 & -73.04695 & 0.97 & 16.57 & 16.38 & 6 & [He/H=0] & DA & $6680^{+64}_{-64}$ & $7.665^{+0.019}_{-0.019}$ & $0.412^{+0.013}_{-0.012}$ \\
WDJ045312.76-442340.14 & 4811421896276768128 & 73.30335 & -44.39438 & 1.00 & 15.45 & 15.55 & 12 & [He/H=0] & DA & $19017^{+450}_{-449}$ & $7.819^{+0.027}_{-0.026}$ & $0.525^{+0.019}_{-0.017}$ \\
WDJ045359.85-432247.75 & 4811871287295222016 & 73.50155 & -43.37959 & 0.99 & 16.82 & 16.82 & 12 & [He/H=0] & DA & $11162^{+196}_{-196}$ & $7.606^{+0.019}_{-0.018}$ & $0.404^{+0.011}_{-0.011}$ \\
WDJ045405.87-342505.71 & 4873109308958466432 & 73.52446 & -34.41821 & 1.00 & 16.13 & 16.15 & 6 & [He/H=0] & DA & $11960^{+153}_{-152}$ & $8.028^{+0.009}_{-0.010}$ & $0.622^{+0.008}_{-0.008}$ \\
WDJ045423.71-343948.35 & 4873054711334208256 & 73.59970 & -34.66404 & 1.00 & 16.00 & 16.11 & 6 & [He/H=0] & DA & $20271^{+269}_{-269}$ & $7.796^{+0.011}_{-0.011}$ & $0.518^{+0.007}_{-0.008}$ \\
WDJ045440.28-342651.96 & 4873061617641604864 & 73.66829 & -34.44714 & 1.00 & 15.98 & 16.08 & 6 & [He/H=0] & DA & $19414^{+287}_{-287}$ & $7.802^{+0.013}_{-0.012}$ & $0.519^{+0.008}_{-0.009}$ \\
WDJ045514.63-544145.41 & 4776747972460494208 & 73.80995 & -54.69619 & 1.00 & 14.37 & 14.45 & 12 & [He/H=0] & DA & $16659^{+215}_{-215}$ & $7.797^{+0.011}_{-0.011}$ & $0.508^{+0.008}_{-0.008}$ \\
WDJ045527.27-625844.61 & 4665237908354588288 & 73.86402 & -62.97854 & 1.00 & 14.97 & 14.97 & 6 & [He/H=0] & DAV & $11106^{+115}_{-115}$ & $7.840^{+0.012}_{-0.011}$ & $0.514^{+0.009}_{-0.008}$ \\
WDJ045535.94-292859.95 & 4876689387538123008 & 73.89973 & -29.48293 & 1.00 & 15.01 & 15.07 & 6 & [He/H=0] & DA & $16230^{+133}_{-134}$ & $7.218^{+0.007}_{-0.007}$ & $0.296^{+0.003}_{-0.002}$ \\
WDJ045551.33-705541.81 & 4654695103690489856 & 73.96579 & -70.92856 & 0.99 & 16.49 & 16.45 & 6 & [He/H=0] & DA & $10324^{+238}_{-237}$ & $7.934^{+0.032}_{-0.030}$ & $0.563^{+0.026}_{-0.024}$ \\
WDJ045648.37-435123.35 & 4811658909752355584 & 74.20160 & -43.85660 & 1.00 & 16.12 & 16.22 & 12 & [He/H=0] & DA & $19649^{+299}_{-298}$ & $7.864^{+0.014}_{-0.014}$ & $0.550^{+0.010}_{-0.010}$ \\
WDJ045658.40-531026.40 & 4782942513595999872 & 74.24372 & -53.17393 & 1.00 & 16.79 & 16.90 & 12 & [He/H=0] & DA & $21331^{+443}_{-443}$ & $7.483^{+0.022}_{-0.021}$ & $0.393^{+0.011}_{-0.009}$ \\
WDJ045713.32-280752.72 & 4880286371109059712 & 74.30576 & -28.13126 & 0.99 & 13.93 & 14.13 & 6 & [He/H=0] & DA & $39575^{+335}_{-335}$ & $7.377^{+0.029}_{-0.034}$ & $0.419^{+0.013}_{-0.014}$ \\
WDJ045732.31-641010.95 & 4664651250184794368 & 74.38533 & -64.16834 & 0.99 & 16.95 & 16.80 & 6 & [He/H=0] & DA & $6293^{+103}_{-104}$ & $7.340^{+0.055}_{-0.050}$ & $0.282^{+0.027}_{-0.022}$ \\
WDJ050042.52-301637.70 & 4876440794831381760 & 75.17721 & -30.27707 & 0.98 & 16.63 & 16.84 & 6 & [He/H=0] & DA & $97759^{+7538}_{-7538}$ & $7.587^{+0.083}_{-0.071}$ & $0.645^{+0.033}_{-0.030}$ \\
WDJ050323.06-570122.22 & 4763928766392708864 & 75.84610 & -57.02269 & 0.99 & 16.11 & 16.30 & 12 & [He/H=0] & DA & $41388^{+3381}_{-3381}$ & $7.393^{+0.051}_{-0.047}$ & $0.430^{+0.025}_{-0.025}$ \\
WDJ050458.06-494143.53 & 4785276918518079616 & 76.24243 & -49.69469 & 1.00 & 15.90 & 15.96 & 12 & [He/H=0] & DA & $15133^{+202}_{-202}$ & $7.888^{+0.013}_{-0.012}$ & $0.551^{+0.010}_{-0.009}$ \\
WDJ051223.04-414526.04 & 4812859061053900928 & 78.09605 & -41.75724 & 1.00 & 16.84 & 17.03 & 18 & [He/H=0] & DA & $45626^{+5163}_{-5162}$ & $7.836^{+0.062}_{-0.056}$ & $0.600^{+0.041}_{-0.036}$ \\
WDJ051327.80-412751.88 & 4812870056170125568 & 78.36593 & -41.46307 & 1.00 & 15.98 & 15.95 & 18 & [He/H=0] & DA & $10122^{+162}_{-161}$ & $7.957^{+0.022}_{-0.022}$ & $0.575^{+0.019}_{-0.017}$ \\
WDJ051537.22-322454.03 & 4827311522925744768 & 78.90534 & -32.41507 & 1.00 & 16.32 & 16.41 & 12 & [He/H=0] & DA & $18110^{+301}_{-301}$ & $7.653^{+0.015}_{-0.015}$ & $0.446^{+0.010}_{-0.008}$ \\
WDJ052042.54-361610.81 & 4822336743909344256 & 80.17721 & -36.26987 & 1.00 & 16.39 & 16.49 & 12 & [He/H=0] & DA & $18504^{+304}_{-303}$ & $7.853^{+0.015}_{-0.015}$ & $0.541^{+0.011}_{-0.010}$ \\
WDJ052422.51-514054.85 & 4772418885944906112 & 81.09419 & -51.68156 & 1.00 & 15.92 & 15.96 & 24 & [He/H=0] & DA & $13328^{+190}_{-190}$ & $7.938^{+0.012}_{-0.012}$ & $0.574^{+0.009}_{-0.010}$ \\
WDJ052446.21-381837.14 & 4819672528450697984 & 81.19251 & -38.31044 & 0.99 & 16.60 & 16.61 & 18 & [He/H=0] & DA & $20887^{+604}_{-604}$ & $7.785^{+0.028}_{-0.027}$ & $0.514^{+0.019}_{-0.017}$ \\
WDJ052448.16-464100.34 & 4798944672685041152 & 81.20107 & -46.68432 & 0.99 & 16.18 & 16.09 & 24 & [He/H=0] & DA & $8578^{+92}_{-92}$ & $7.873^{+0.017}_{-0.017}$ & $0.524^{+0.013}_{-0.013}$ \\
WDJ052518.96-302018.13 & 2905372374461903488 & 81.32921 & -30.33830 & 0.98 & 16.76 & 16.77 & 12 & [He/H=0] & DA & $11616^{+109}_{-109}$ & $7.904^{+0.010}_{-0.010}$ & $0.550^{+0.008}_{-0.008}$ \\
WDJ052528.17-385410.97 & 4807583638623283584 & 81.36772 & -38.90262 & 0.04 & 16.23 & 16.04 & 18 & [He/H=0] & DAM & $16654^{+326}_{-325}$ & $7.319^{+0.081}_{-0.097}$ & $0.325^{+0.033}_{-0.035}$ \\
WDJ052545.71-544407.34 & 4769992263782028544 & 81.44031 & -54.73556 & 1.00 & 16.47 & 16.61 & 18 & [He/H=0] & DA & $26078^{+465}_{-464}$ & $7.674^{+0.020}_{-0.021}$ & $0.480^{+0.012}_{-0.011}$ \\
WDJ052548.76-671611.58 & 4658691072554066560 & 81.45281 & -67.27038 & 1.00 & 16.91 & 16.86 & 6 & [He/H=0] & DA & $10493^{+328}_{-328}$ & $8.127^{+0.047}_{-0.043}$ & $0.679^{+0.041}_{-0.038}$ \\
WDJ052639.42-353557.43 & 4822043831436262784 & 81.66460 & -35.59954 & 0.97 & 16.17 & 16.12 & 12 & [He/H=0] & DA & $9522^{+78}_{-78}$ & $7.304^{+0.015}_{-0.015}$ & $0.288^{+0.006}_{-0.007}$ \\
WDJ052724.33-310655.56 & 2905067333001582720 & 81.85065 & -31.11690 & 1.00 & 15.96 & 15.99 & 12 & [He/H=0] & DA & $13539^{+238}_{-237}$ & $8.853^{+0.010}_{-0.010}$ & $1.130^{+0.006}_{-0.006}$ \\
WDJ052844.01-430449.21 & 4805691447831511680 & 82.18385 & -43.08050 & 0.99 & 16.09 & 16.04 & 12 & [He/H=0] & DA & $10149^{+124}_{-123}$ & $8.627^{+0.015}_{-0.015}$ & $1.000^{+0.012}_{-0.012}$ \\
WDJ052912.10-430334.49 & 4805782462481529600 & 82.30017 & -43.05951 & 1.00 & 16.17 & 16.19 & 12 & [He/H=0] & DA & $11669^{+180}_{-180}$ & $7.715^{+0.014}_{-0.013}$ & $0.454^{+0.009}_{-0.009}$ \\
WDJ053100.49-455801.14 & 4799224635833122304 & 82.75279 & -45.96644 & 1.00 & 16.66 & 16.63 & 24 & [He/H=0] & DA & $10387^{+184}_{-184}$ & $7.948^{+0.025}_{-0.025}$ & $0.571^{+0.021}_{-0.020}$ \\
WDJ053203.91-653609.91 & 4660769493122896128 & 83.01610 & -65.60277 & 1.00 & 16.89 & 16.89 & 12 & [He/H=0] & DA & $11330^{+250}_{-250}$ & $7.904^{+0.025}_{-0.025}$ & $0.549^{+0.020}_{-0.019}$ \\
WDJ053306.78-560353.38 & 4766810380210581632 & 83.27841 & -56.06476 & 1.00 & 15.93 & 15.92 & 18 & [He/H=0] & DA & $10915^{+121}_{-121}$ & $8.345^{+0.014}_{-0.013}$ & $0.820^{+0.012}_{-0.011}$ \\
WDJ053343.43-271350.08 & 2908425653829891712 & 83.43077 & -27.23082 & 1.00 & 15.61 & 15.73 & 12 & [He/H=0] & DA & $22113^{+235}_{-235}$ & $9.004^{+0.007}_{-0.007}$ & $1.204^{+0.004}_{-0.003}$ \\
WDJ053640.24-325454.63 & 2901629916055849472 & 84.16764 & -32.91517 & 1.00 & 16.82 & 16.92 & 12 & [He/H=0] & DA & $19138^{+509}_{-509}$ & $7.843^{+0.023}_{-0.022}$ & $0.538^{+0.016}_{-0.015}$ \\
WDJ053753.46-475805.10 & 4795556287084999552 & 84.47255 & -47.96814 & 1.00 & 15.59 & 15.71 & 24 & [He/H=0] & DA & $21643^{+390}_{-391}$ & $8.049^{+0.016}_{-0.015}$ & $0.656^{+0.012}_{-0.012}$ \\
WDJ053939.69-664157.85 & 4659528969150132096 & 84.91658 & -66.69852 & 1.00 & 15.98 & 15.99 & 12 & [He/H=0] & DA & $11868^{+263}_{-264}$ & $7.912^{+0.021}_{-0.021}$ & $0.555^{+0.017}_{-0.016}$ \\
WDJ054412.15-352547.00 & 2888177253850677760 & 86.05097 & -35.43001 & 1.00 & 15.23 & 15.25 & 12 & [He/H=0] & DA & $12573^{+186}_{-185}$ & $7.888^{+0.010}_{-0.010}$ & $0.544^{+0.008}_{-0.008}$ \\
WDJ054458.33-401940.13 & 4805139866655106944 & 86.24291 & -40.32770 & 1.00 & 16.96 & 16.95 & 24 & [He/H=0] & DA & $11596^{+268}_{-269}$ & $7.942^{+0.025}_{-0.024}$ & $0.571^{+0.020}_{-0.019}$ \\
WDJ055428.05-402555.11 & 4804209684114310784 & 88.61677 & -40.43208 & 1.00 & 16.63 & 16.65 & 24 & [He/H=0] & DA & $12570^{+338}_{-337}$ & $8.354^{+0.021}_{-0.020}$ & $0.829^{+0.018}_{-0.018}$ \\
WDJ055814.63-373426.02 & 2884285498084601600 & 89.56110 & -37.57417 & 0.99 & 14.32 & 14.53 & 24 & [He/H=0] & DA & $54126^{+4471}_{-4471}$ & $7.309^{+0.041}_{-0.038}$ & $0.434^{+0.025}_{-0.021}$ \\
WDJ060013.07-545516.04 & 5500308124837433984 & 90.05431 & -54.92130 & 1.00 & 16.21 & 16.21 & 18 & [He/H=0] & DA & $10785^{+139}_{-139}$ & $8.202^{+0.016}_{-0.016}$ & $0.727^{+0.014}_{-0.014}$ \\
WDJ060502.73-481958.44 & 5554202096021588992 & 91.26148 & -48.33261 & 1.00 & 15.83 & 16.00 & 24 & [He/H=0] & DA & $31780^{+705}_{-705}$ & $7.702^{+0.023}_{-0.022}$ & $0.508^{+0.013}_{-0.013}$ \\
WDJ060843.80-530134.27 & 5548648634588556288 & 92.18144 & -53.02523 & 0.99 & 15.96 & 15.90 & 24 & [He/H=0] & DA & $9560^{+98}_{-98}$ & $7.887^{+0.017}_{-0.016}$ & $0.535^{+0.012}_{-0.013}$ \\
WDJ061958.87-414337.84 & 5571715735704543360 & 94.99655 & -41.72883 & 1.00 & 15.41 & 15.35 & 24 & [He/H=0] & DA & $9450^{+86}_{-86}$ & $7.933^{+0.015}_{-0.014}$ & $0.560^{+0.012}_{-0.011}$ \\
WDJ062312.62-374128.04 & 5575007845317435648 & 95.80295 & -37.69113 & 0.98 & 12.02 & 12.23 & 24 & [He/H=0] & DA & $50596^{+4372}_{-4371}$ & $7.120^{+0.045}_{-0.042}$ & $0.375^{+0.024}_{-0.020}$ \\
WDJ062425.79-325727.38 & 2892228919840595072 & 96.10746 & -32.95841 & 0.99 & 15.42 & 15.40 & 12 & [He/H=0] & DAB & $13614^{+142}_{-141}$ & $7.733^{+0.010}_{-0.009}$ & $0.469^{+0.006}_{-0.006}$ \\
WDJ064428.73-283238.07 & 2918665577420228480 & 101.11914 & -28.54434 & 0.99 & 16.42 & 16.34 & 12 & [He/H=0] & DA & $8873^{+54}_{-54}$ & $7.839^{+0.011}_{-0.010}$ & $0.507^{+0.008}_{-0.008}$ \\
WDJ064802.46-483039.16 & 5551402808137892864 & 102.01025 & -48.51111 & 1.00 & 16.73 & 16.74 & 24 & [He/H=0] & DA & $11551^{+201}_{-201}$ & $7.920^{+0.019}_{-0.019}$ & $0.559^{+0.015}_{-0.015}$ \\
WDJ064856.09-252346.96 & 2921786919133198848 & 102.23357 & -25.39623 & 1.00 & 13.63 & 13.78 & 6 & [He/H=0] & DA & $27028^{+213}_{-213}$ & $7.869^{+0.008}_{-0.008}$ & $0.571^{+0.006}_{-0.005}$ \\
WDJ065330.21-395429.12 & 5564029702750970112 & 103.37551 & -39.90710 & 0.99 & 15.92 & 15.70 & 24 & [He/H=0] & DA & $6177^{+27}_{-27}$ & $8.046^{+0.009}_{-0.010}$ & $0.617^{+0.008}_{-0.009}$ \\
WDJ065335.34-395533.29 & 5564028981196462336 & 103.39689 & -39.92490 & 0.99 & 15.35 & 15.19 & 24 & [He/H=0] & DA & $6975^{+38}_{-39}$ & $7.965^{+0.011}_{-0.011}$ & $0.571^{+0.009}_{-0.009}$ \\
WDJ065355.03-562447.16 & 5484935887129131648 & 103.47934 & -56.41298 & 1.00 & 16.62 & 16.79 & 24 & [He/H=0] & DA & $30025^{+701}_{-701}$ & $8.719^{+0.022}_{-0.021}$ & $1.071^{+0.014}_{-0.015}$ \\
WDJ065705.91-390935.68 & 5564171814627287296 & 104.27292 & -39.16058 & 0.99 & 14.98 & 14.77 & 6 & [He/H=0] & DA & $6356^{+26}_{-26}$ & $8.068^{+0.009}_{-0.009}$ & $0.631^{+0.008}_{-0.008}$ \\
WDJ070219.72-585009.02 & 5480556532316697216 & 105.58175 & -58.83494 & 1.00 & 14.58 & 14.61 & 12 & [He/H=0] & DA & $12929^{+179}_{-180}$ & $8.435^{+0.009}_{-0.009}$ & $0.882^{+0.007}_{-0.007}$ \\
WDJ070520.53-204642.21 & 2929364817693200640 & 106.33599 & -20.77873 & 1.00 & 16.50 & 16.67 & 6 & [He/H=0] & DAH & $29476^{+402}_{-402}$ & $8.880^{+0.012}_{-0.013}$ & $1.153^{+0.007}_{-0.007}$ \\
WDJ070925.08-320507.33 & 5604216131195297664 & 107.35349 & -32.08313 & 1.00 & 15.60 & 15.55 & 6 & [He/H=0] & DA & $9736^{+149}_{-149}$ & $7.915^{+0.023}_{-0.023}$ & $0.550^{+0.019}_{-0.017}$ \\
WDJ071039.41-414424.56 & 5561160355360035456 & 107.66520 & -41.74069 & 1.00 & 15.79 & 15.77 & 24 & [He/H=0] & DAH & $10074^{+91}_{-90}$ & $7.933^{+0.014}_{-0.013}$ & $0.562^{+0.011}_{-0.011}$ \\
WDJ071051.17-342107.14 & 5602672550010235008 & 107.71328 & -34.35244 & 1.00 & 16.24 & 16.22 & 12 & [He/H=0] & DA & $10707^{+127}_{-126}$ & $7.968^{+0.014}_{-0.013}$ & $0.584^{+0.011}_{-0.011}$ \\
WDJ071114.41-251815.05 & 5617205933369326592 & 107.80948 & -25.30330 & 0.99 & 14.45 & 14.42 & 6 & [He/H=0] & DA & $10677^{+39}_{-40}$ & $7.141^{+0.004}_{-0.004}$ & $0.251^{+0.001}_{-0.001}$ \\
WDJ071516.58-702505.62 & 5267418883330985856 & 108.81764 & -70.41772 & 1.00 & 14.01 & 14.20 & 12 & [He/H=0] & DA & $37834^{+1551}_{-1552}$ & $7.737^{+0.026}_{-0.025}$ & $0.538^{+0.016}_{-0.014}$ \\
WDJ071550.55-370642.20 & 5589354543620145024 & 108.95916 & -37.11102 & 1.00 & 16.59 & 16.43 & 12 & [He/H=0] & DA & $7195^{+59}_{-60}$ & $8.399^{+0.013}_{-0.014}$ & $0.849^{+0.012}_{-0.012}$ \\
WDJ071751.88-282928.84 & 5606764386835758720 & 109.46577 & -28.49113 & 1.00 & 16.42 & 16.39 & 6 & [He/H=0] & DA & $10444^{+50}_{-50}$ & $7.988^{+0.008}_{-0.007}$ & $0.595^{+0.006}_{-0.007}$ \\
WDJ072047.91-314702.73 & 5604890028746727552 & 110.19944 & -31.78405 & 0.06 & 14.57 & 14.45 & 6 & [He/H=0] & DA & $28292^{+654}_{-654}$ & $6.918^{+0.027}_{-0.026}$ & $0.261^{+0.008}_{-0.008}$ \\
WDJ072320.10-274721.58 & 5612710511362093184 & 110.83391 & -27.78942 & 1.00 & 14.52 & 14.69 & 6 & [He/H=0] & DA & $33637^{+699}_{-699}$ & $7.765^{+0.017}_{-0.018}$ & $0.539^{+0.011}_{-0.010}$ \\
WDJ072717.48-332838.16 & 5592485716514387584 & 111.82298 & -33.47756 & 1.00 & 16.55 & 16.55 & 12 & [He/H=0] & DA & $11775^{+297}_{-298}$ & $7.714^{+0.023}_{-0.022}$ & $0.454^{+0.016}_{-0.015}$ \\
WDJ073326.40-445325.34 & 5510964144860346496 & 113.36057 & -44.89046 & 0.99 & 15.34 & 15.26 & 12 & [He/H=0] & DA & $9181^{+82}_{-82}$ & $7.937^{+0.014}_{-0.015}$ & $0.561^{+0.012}_{-0.011}$ \\
WDJ073337.81-425357.25 & 5536077746353130240 & 113.40796 & -42.89623 & 1.00 & 14.19 & 14.24 & 12 & [He/H=0] & DA & $14726^{+223}_{-223}$ & $8.111^{+0.014}_{-0.014}$ & $0.678^{+0.012}_{-0.012}$ \\
WDJ073338.28-281623.17 & 5611744006283630720 & 113.40918 & -28.27263 & 1.00 & 15.96 & 15.97 & 6 & [He/H=0] & DA & $12351^{+62}_{-62}$ & $8.373^{+0.004}_{-0.005}$ & $0.841^{+0.004}_{-0.004}$ \\
WDJ074152.84-570844.74 & 5295512298775133824 & 115.46997 & -57.14541 & 1.00 & 15.15 & 15.25 & 12 & [He/H=0] & DA & $19549^{+311}_{-311}$ & $7.668^{+0.014}_{-0.014}$ & $0.458^{+0.008}_{-0.008}$ \\
WDJ074613.81-302550.93 & 5598661329740179712 & 116.55786 & -30.43134 & 1.00 & 16.03 & 16.01 & 6 & [He/H=0] & DA & $10754^{+172}_{-172}$ & $7.923^{+0.018}_{-0.018}$ & $0.558^{+0.014}_{-0.014}$ \\
WDJ075308.14-674731.38 & 5273943488410008832 & 118.30118 & -67.79867 & 1.00 & 13.78 & 13.53 & 12 & [He/H=0] & DA & $5636^{+12}_{-12}$ & $7.969^{+0.006}_{-0.005}$ & $0.567^{+0.005}_{-0.004}$ \\
WDJ075328.47-511436.98 & 5513896164414899456 & 118.36809 & -51.24283 & 0.99 & 15.61 & 15.53 & 12 & [He/H=0] & DAH & $9117^{+81}_{-81}$ & $8.343^{+0.013}_{-0.013}$ & $0.815^{+0.012}_{-0.011}$ \\
WDJ075643.91-545348.89 & 5296195267296378112 & 119.18267 & -54.89675 & 1.00 & 15.52 & 15.57 & 12 & [He/H=0] & DA & $11549^{+445}_{-445}$ & $7.871^{+0.046}_{-0.044}$ & $0.532^{+0.036}_{-0.032}$ \\
WDJ080200.19-532750.82 & 5512466799294681472 & 120.50156 & -53.46395 & 0.08 & 15.82 & 15.64 & 6 & [He/H=0] & DA & $16937^{+311}_{-311}$ & $7.573^{+0.019}_{-0.019}$ & $0.410^{+0.011}_{-0.010}$ \\
WDJ081227.07-352943.32 & 5544743925212648320 & 123.11242 & -35.49550 & 1.00 & 14.35 & 14.14 & 12 & [He/H=0] & DAH & $6238^{+24}_{-24}$ & $8.179^{+0.008}_{-0.008}$ & $0.701^{+0.008}_{-0.007}$ \\
WDJ082126.70-670320.02 & 5272690766709543680 & 125.35673 & -67.05263 & 1.00 & 15.08 & 14.76 & 6 & [He/H=0] & DA & $4807^{+39}_{-40}$ & $7.937^{+0.020}_{-0.020}$ & $0.541^{+0.017}_{-0.016}$ \\
WDJ083151.91-534032.52 & 5321056462156468480 & 127.96650 & -53.67625 & 1.00 & 14.45 & 14.62 & 6 & [He/H=0] & DA & $28395^{+532}_{-532}$ & $7.810^{+0.020}_{-0.018}$ & $0.546^{+0.013}_{-0.012}$ \\
WDJ083246.58-574125.49 & 5315128548292806656 & 128.19400 & -57.69018 & 1.00 & 16.69 & 16.69 & 6 & [He/H=0] & DA & $11812^{+222}_{-223}$ & $8.039^{+0.019}_{-0.019}$ & $0.628^{+0.016}_{-0.016}$ \\
WDJ083759.16-501745.76 & 5322090003089341440 & 129.49465 & -50.29528 & 1.00 & 14.60 & 14.61 & 6 & [He/H=0] & DA & $12163^{+194}_{-194}$ & $8.259^{+0.013}_{-0.013}$ & $0.767^{+0.011}_{-0.012}$ \\
WDJ084215.26-555628.51 & 5316824544982104960 & 130.56328 & -55.94139 & 1.00 & 16.52 & 16.49 & 6 & [He/H=0] & DA & $10358^{+181}_{-182}$ & $7.933^{+0.026}_{-0.025}$ & $0.563^{+0.020}_{-0.020}$ \\
WDJ085102.69-615517.65 & 5298297567892486656 & 132.76058 & -61.92144 & 1.00 & 14.71 & 14.82 & 6 & [He/H=0] & DA & $19665^{+266}_{-266}$ & $7.961^{+0.012}_{-0.012}$ & $0.601^{+0.010}_{-0.009}$ \\

\label{tab:DA}
\end{longtable}

\end{landscape}


\onecolumn
\begin{landscape}

\small\tabcolsep=1.5pt

\centering

\begin{longtable}{ccccccccccccc}

\caption{Parameters of white dwarfs modeled with mixed H/He atmospheres are presented, incorporating Gaia DR3 data, PLATO Input Catalog information, and values derived from our photometric fits, with targets ordered by increasing right ascension. } \\

\toprule
Name & Gaia ID & RA [deg] & Dec [deg] & Pwd & $G$ [mag] & $P$ [mag] & $n_{\rm CAM}$ & Abundance & S. Type &  $T_{\rm eff}$ [K] & $\log g$ [cm\,s$^{-2}$] & $M$ [$M_\odot$] \\
\midrule
\endfirsthead

\toprule
Name & Gaia ID & RA [deg] & Dec [deg] & Pwd & $G$ [mag] & $P$ [mag] & $n_{\rm CAM}$ & Abundance & S. Type &  $T_{\rm eff}$ [K] & $\log g$ [cm\,s$^{-2}$] & $M$ [$M_\odot$] \\
\midrule
\endhead
\midrule
\multicolumn{13}{r}{{Continued on next page}} \\
\midrule
\endfoot

\bottomrule
\endlastfoot

WDJ033117.34-592537.38 & 4729070979320996096 & 52.82229 & -59.42707 & 1.00 & 16.25 & 16.42 & 6 & [log H/He=-5.00] & DB & $39107^{+1019}_{-1020}$ & $7.801^{+0.026}_{-0.024}$ & $0.535^{+0.015}_{-0.015}$ \\
WDJ041909.91-535045.31 & 4779379172505736832 & 64.79099 & -53.84642 & 1.00 & 15.27 & 15.36 & 12 & [log H/He=-5.00] & DB & $17617^{+496}_{-497}$ & $7.871^{+0.028}_{-0.027}$ & $0.524^{+0.021}_{-0.020}$ \\
WDJ042211.36-474142.02 & 4788896304637763072 & 65.54690 & -47.69497 & 1.00 & 15.23 & 15.34 & 12 & [log H/He=-5.00] & DB & $19960^{+515}_{-515}$ & $7.830^{+0.020}_{-0.019}$ & $0.508^{+0.014}_{-0.013}$ \\
WDJ042432.05-545539.23 & 4778990563864168448 & 66.13233 & -54.92781 & 1.00 & 16.91 & 16.95 & 12 & [log H/He=-5.00] & DB & $12985^{+267}_{-267}$ & $7.876^{+0.028}_{-0.026}$ & $0.517^{+0.022}_{-0.019}$ \\
WDJ050007.86-600728.79 & 4761685522153733760 & 75.03280 & -60.12458 & 1.00 & 16.29 & 16.34 & 12 & [log H/He=-5.00] & DB & $14181^{+277}_{-277}$ & $7.901^{+0.023}_{-0.023}$ & $0.533^{+0.017}_{-0.017}$ \\
WDJ061614.26-591227.41 & 5482551252566796928 & 94.05898 & -59.20900 & 1.00 & 13.95 & 14.02 & 24 & [log H/He=-5.00] & DB & $14336^{+303}_{-303}$ & $7.859^{+0.027}_{-0.025}$ & $0.511^{+0.020}_{-0.018}$ \\
WDJ044427.35-351745.09 & 4867104257483687040 & 71.11383 & -35.29618 & 1.00 & 15.68 & 15.71 & 6 & [log H/He=-5.00] & DBA & $12793^{+179}_{-180}$ & $8.125^{+0.017}_{-0.017}$ & $0.662^{+0.015}_{-0.015}$ \\
WDJ045153.75-254914.73 & 4881672923991046400 & 72.97381 & -25.82079 & 1.00 & 16.43 & 16.41 & 6 & [log H/He=-5.00] & DBAZ & $10237^{+28}_{-28}$ & $8.038^{+0.005}_{-0.006}$ & $0.604^{+0.004}_{-0.005}$ \\
WDJ052330.80-472219.55 & 4797409582653745152 & 80.87843 & -47.37214 & 1.00 & 16.69 & 16.80 & 24 & [log H/He=-5.00] & DBV & $19045^{+766}_{-765}$ & $7.858^{+0.037}_{-0.034}$ & $0.520^{+0.027}_{-0.024}$ \\
WDJ041724.58-673236.77 & 4668242152077665024 & 64.35251 & -67.54344 & 1.00 & 16.63 & 16.75 & 6 & [log H/He=-5.00] & DB & $17419^{+875}_{-874}$ & $7.671^{+0.053}_{-0.049}$ & $0.428^{+0.034}_{-0.029}$ \\
WDJ042610.85-465847.47 & 4789144175789941888 & 66.54522 & -46.98000 & 1.00 & 16.78 & 16.91 & 12 & [log H/He=-5.00] & DB & $20442^{+819}_{-819}$ & $7.831^{+0.028}_{-0.027}$ & $0.510^{+0.019}_{-0.019}$ \\
WDJ043151.06-532929.85 & 4780712394777586560 & 67.96298 & -53.49164 & 1.00 & 16.98 & 17.13 & 12 & [log H/He=-5.00] & DB & $28181^{+1218}_{-1217}$ & $7.806^{+0.035}_{-0.032}$ & $0.514^{+0.023}_{-0.020}$ \\
WDJ050510.82-613924.40 & 4760710942535282560 & 76.29501 & -61.65674 & 1.00 & 16.91 & 17.03 & 6 & [log H/He=-5.00] & DB & $21235^{+713}_{-714}$ & $8.181^{+0.023}_{-0.022}$ & $0.709^{+0.019}_{-0.020}$ \\
WDJ053423.94-381444.86 & 4808986478021189120 & 83.59984 & -38.24568 & 1.00 & 16.91 & 17.05 & 12 & [log H/He=-5.00] & DB & $26509^{+1437}_{-1437}$ & $7.536^{+0.038}_{-0.036}$ & $0.401^{+0.019}_{-0.017}$ \\
WDJ054007.06-651807.95 & 4660085867450146560 & 85.02894 & -65.30190 & 1.00 & 16.50 & 16.60 & 12 & [log H/He=-5.00] & DB & $17084^{+1291}_{-1291}$ & $7.739^{+0.091}_{-0.078}$ & $0.458^{+0.062}_{-0.048}$ \\
WDJ054657.41-305153.89 & 2903068214703934848 & 86.73922 & -30.86496 & 1.00 & 16.51 & 16.64 & 12 & [log H/He=-5.00] & DB & $25266^{+1650}_{-1649}$ & $8.226^{+0.019}_{-0.018}$ & $0.742^{+0.017}_{-0.015}$ \\
WDJ055635.50-561006.57 & 4767225652008399104 & 89.14807 & -56.16833 & 1.00 & 15.44 & 15.56 & 24 & [log H/He=-5.00] & DB & $19561^{+489}_{-489}$ & $8.232^{+0.020}_{-0.019}$ & $0.739^{+0.017}_{-0.017}$ \\
WDJ060612.38-723414.20 & 5266145236552677632 & 91.55152 & -72.57079 & 1.00 & 16.67 & 16.77 & 6 & [log H/He=-5.00] & DB & $18259^{+742}_{-742}$ & $7.453^{+0.038}_{-0.035}$ & $0.348^{+0.018}_{-0.015}$ \\
WDJ060740.96-285302.51 & 2897810350100517504 & 91.92038 & -28.88393 & 1.00 & 16.27 & 16.39 & 12 & [log H/He=-5.00] & DB & $18283^{+264}_{-265}$ & $7.789^{+0.014}_{-0.013}$ & $0.484^{+0.010}_{-0.009}$ \\
WDJ061109.12-703921.89 & 5278615205944625664 & 92.78797 & -70.65557 & 1.00 & 16.87 & 16.99 & 6 & [log H/He=-5.00] & DB & $26597^{+1411}_{-1411}$ & $7.760^{+0.042}_{-0.039}$ & $0.490^{+0.027}_{-0.024}$ \\
WDJ061238.45-325357.56 & 2892795821161588736 & 93.16004 & -32.89928 & 1.00 & 16.23 & 16.34 & 12 & [log H/He=-5.00] & DB & $19180^{+484}_{-484}$ & $7.657^{+0.022}_{-0.021}$ & $0.427^{+0.013}_{-0.013}$ \\
WDJ061524.39-384823.63 & 5574654180530308736 & 93.85156 & -38.80646 & 1.00 & 16.55 & 16.68 & 24 & [log H/He=-5.00] & DB & $25422^{+827}_{-826}$ & $7.768^{+0.020}_{-0.019}$ & $0.491^{+0.012}_{-0.013}$ \\
WDJ061856.82-372607.54 & 5575200775248718976 & 94.73673 & -37.43540 & 1.00 & 16.45 & 16.58 & 24 & [log H/He=-5.00] & DB & $22594^{+1746}_{-1746}$ & $7.787^{+0.022}_{-0.020}$ & $0.493^{+0.015}_{-0.014}$ \\
WDJ062000.90-233500.19 & 2936643760765256576 & 95.00382 & -23.58340 & 1.00 & 16.64 & 16.71 & 6 & [log H/He=-5.00] & DB & $21047^{+603}_{-604}$ & $7.682^{+0.022}_{-0.022}$ & $0.442^{+0.014}_{-0.013}$ \\
WDJ063814.10-694335.39 & 5278903110486861312 & 99.55865 & -69.72632 & 1.00 & 16.79 & 16.95 & 12 & [log H/He=-5.00] & DB & $32439^{+1115}_{-1115}$ & $8.047^{+0.029}_{-0.028}$ & $0.646^{+0.023}_{-0.021}$ \\
WDJ063845.14-470141.00 & 5552665081846145024 & 99.68799 & -47.02795 & 1.00 & 16.76 & 16.86 & 24 & [log H/He=-5.00] & DB & $17512^{+775}_{-775}$ & $7.792^{+0.048}_{-0.044}$ & $0.484^{+0.033}_{-0.030}$ \\
WDJ064648.90-525349.76 & 5498647381243346176 & 101.70365 & -52.89709 & 1.00 & 15.91 & 16.03 & 24 & [log H/He=-5.00] & DB & $18988^{+633}_{-633}$ & $7.930^{+0.030}_{-0.028}$ & $0.558^{+0.023}_{-0.021}$ \\
WDJ064747.51-270813.49 & 2919284671187233280 & 101.94803 & -27.13727 & 1.00 & 16.79 & 16.92 & 6 & [log H/He=-5.00] & DB & $28620^{+581}_{-582}$ & $7.893^{+0.018}_{-0.019}$ & $0.557^{+0.013}_{-0.013}$ \\
WDJ065112.06-440937.57 & 5562002134589548032 & 102.80011 & -44.16025 & 1.00 & 16.62 & 16.74 & 24 & [log H/He=-5.00] & DB & $18778^{+899}_{-899}$ & $7.625^{+0.044}_{-0.040}$ & $0.413^{+0.026}_{-0.023}$ \\
WDJ065703.65-433414.37 & 5559127976835354624 & 104.26542 & -43.57085 & 1.00 & 16.41 & 16.54 & 24 & [log H/He=-5.00] & DB & $18403^{+841}_{-841}$ & $7.760^{+0.045}_{-0.041}$ & $0.471^{+0.030}_{-0.027}$ \\
WDJ065957.09-291701.35 & 5608058473357824128 & 104.98775 & -29.28362 & 0.91 & 16.09 & 16.04 & 6 & [log H/He=-5.00] & DB & $21363^{+1236}_{-1237}$ & $7.574^{+0.022}_{-0.022}$ & $0.400^{+0.012}_{-0.012}$ \\
WDJ070918.12-302728.87 & 5604654870693369984 & 107.32535 & -30.45811 & 1.00 & 16.33 & 16.48 & 6 & [log H/He=-5.00] & DB & $29970^{+642}_{-643}$ & $7.723^{+0.017}_{-0.016}$ & $0.481^{+0.011}_{-0.009}$ \\
WDJ073920.28-262915.39 & 5613838438495596032 & 114.83438 & -26.48768 & 1.00 & 16.25 & 16.37 & 6 & [log H/He=-5.00] & DB & $19624^{+303}_{-303}$ & $7.876^{+0.015}_{-0.014}$ & $0.530^{+0.011}_{-0.010}$ \\
WDJ074250.12-544920.12 & 5488182813687216000 & 115.70893 & -54.82214 & 1.00 & 16.90 & 16.99 & 18 & [log H/He=-5.00] & DB & $19127^{+1003}_{-1002}$ & $8.005^{+0.047}_{-0.044}$ & $0.600^{+0.038}_{-0.034}$ \\
WDJ074521.91-432803.38 & 5532479697634251776 & 116.34119 & -43.46757 & 1.00 & 16.75 & 16.86 & 12 & [log H/He=-5.00] & DB & $19536^{+736}_{-735}$ & $8.289^{+0.031}_{-0.030}$ & $0.775^{+0.027}_{-0.026}$ \\
WDJ081612.57-402513.53 & 5539643909237818880 & 124.05225 & -40.42045 & 1.00 & 16.08 & 16.21 & 12 & [log H/He=-5.00] & DB & $21792^{+1440}_{-1441}$ & $7.785^{+0.025}_{-0.024}$ & $0.490^{+0.018}_{-0.016}$ \\
WDJ081827.11-460309.83 & 5519761028164909568 & 124.61303 & -46.05263 & 1.00 & 16.97 & 17.06 & 12 & [log H/He=-5.00] & DB & $17228^{+988}_{-987}$ & $7.843^{+0.062}_{-0.057}$ & $0.508^{+0.046}_{-0.039}$ \\
WDJ082109.21-500114.40 & 5514870018179554560 & 125.28835 & -50.02063 & 0.99 & 16.47 & 16.62 & 12 & [log H/He=-5.00] & DB & $31577^{+977}_{-978}$ & $7.314^{+0.030}_{-0.031}$ & $0.343^{+0.014}_{-0.015}$ \\
WDJ083341.91-655530.74 & 5272852979034659072 & 128.42466 & -65.92519 & 1.00 & 16.71 & 16.83 & 6 & [log H/He=-5.00] & DB & $19752^{+597}_{-596}$ & $7.908^{+0.025}_{-0.024}$ & $0.547^{+0.019}_{-0.017}$ \\
WDJ083942.55-535949.14 & 5317871245692348032 & 129.92690 & -53.99688 & 1.00 & 16.99 & 17.10 & 6 & [log H/He=-5.00] & DB & $20988^{+1040}_{-1039}$ & $7.882^{+0.032}_{-0.032}$ & $0.536^{+0.024}_{-0.022}$ \\
WDJ085424.48-491724.58 & 5328333889104483712 & 133.60189 & -49.29007 & 1.00 & 15.77 & 15.92 & 6 & [log H/He=-5.00] & DB & $30045^{+977}_{-977}$ & $7.901^{+0.025}_{-0.024}$ & $0.564^{+0.018}_{-0.017}$ \\

\label{tab:DB}
\end{longtable}

\end{landscape}

\section{Amplitude spectra from {\tt PlatoSim}}\label{app:amplitude_spectra}

\begin{figure*}[h!]
\center
\includegraphics[width=1\columnwidth]{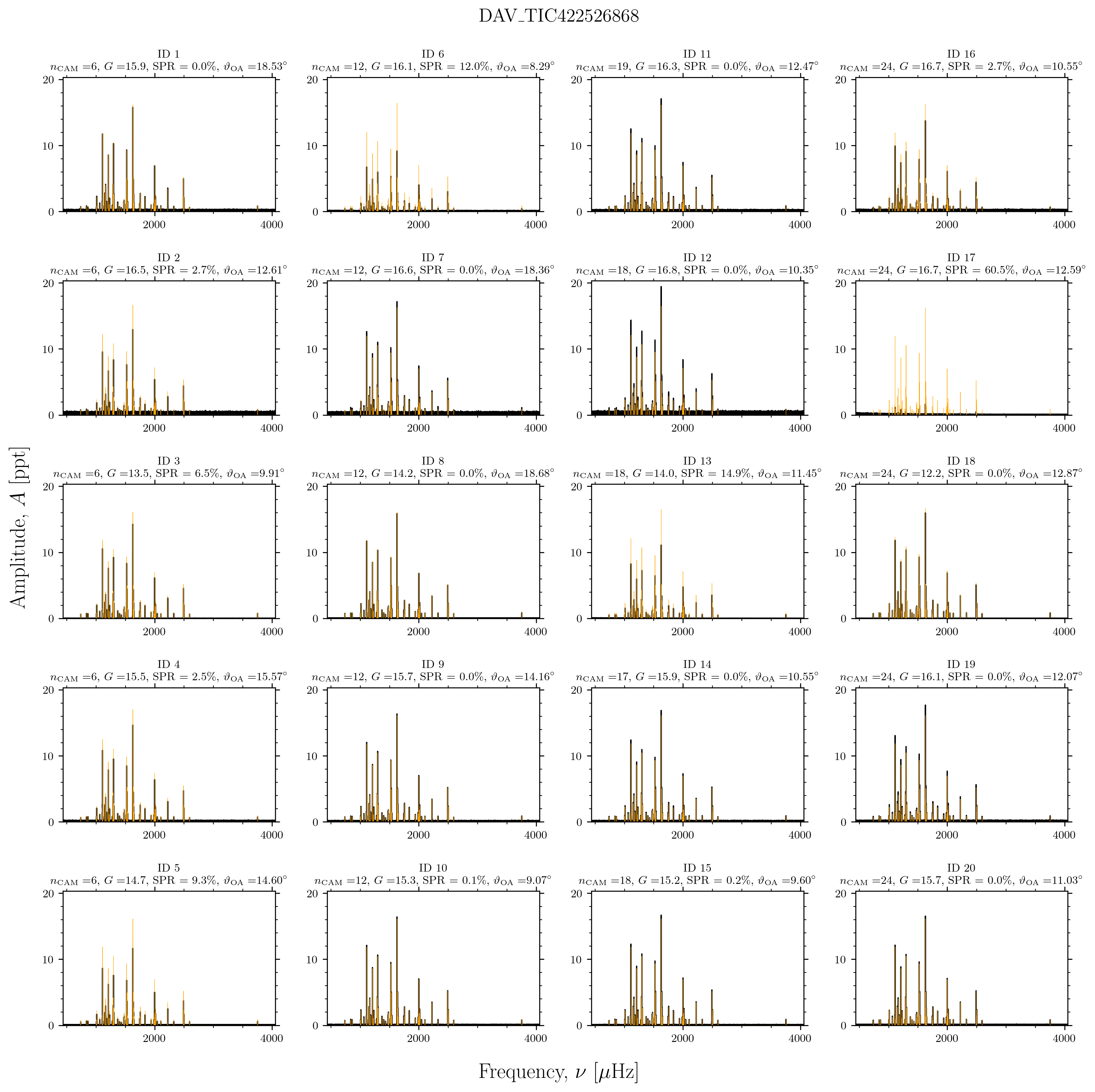}
\caption{Amplitude spectra of the DAV star TIC\,422526868 (G29-38). Each subplot represents one of the 20 mock WD stars (with increasing ID from upper-left to lower-right panel), and  shows the (fast) Lomb-Scargle amplitude spectrum of the simulated (black lines) and injected (orange lines) light curve. The title caption above each panel details information about the mock WD star simulated.} 
\label{fig:amplitude_spectrum_DAV_TIC422526868}
\end{figure*}

\begin{figure*}[h!]
\center
\includegraphics[width=1\columnwidth]{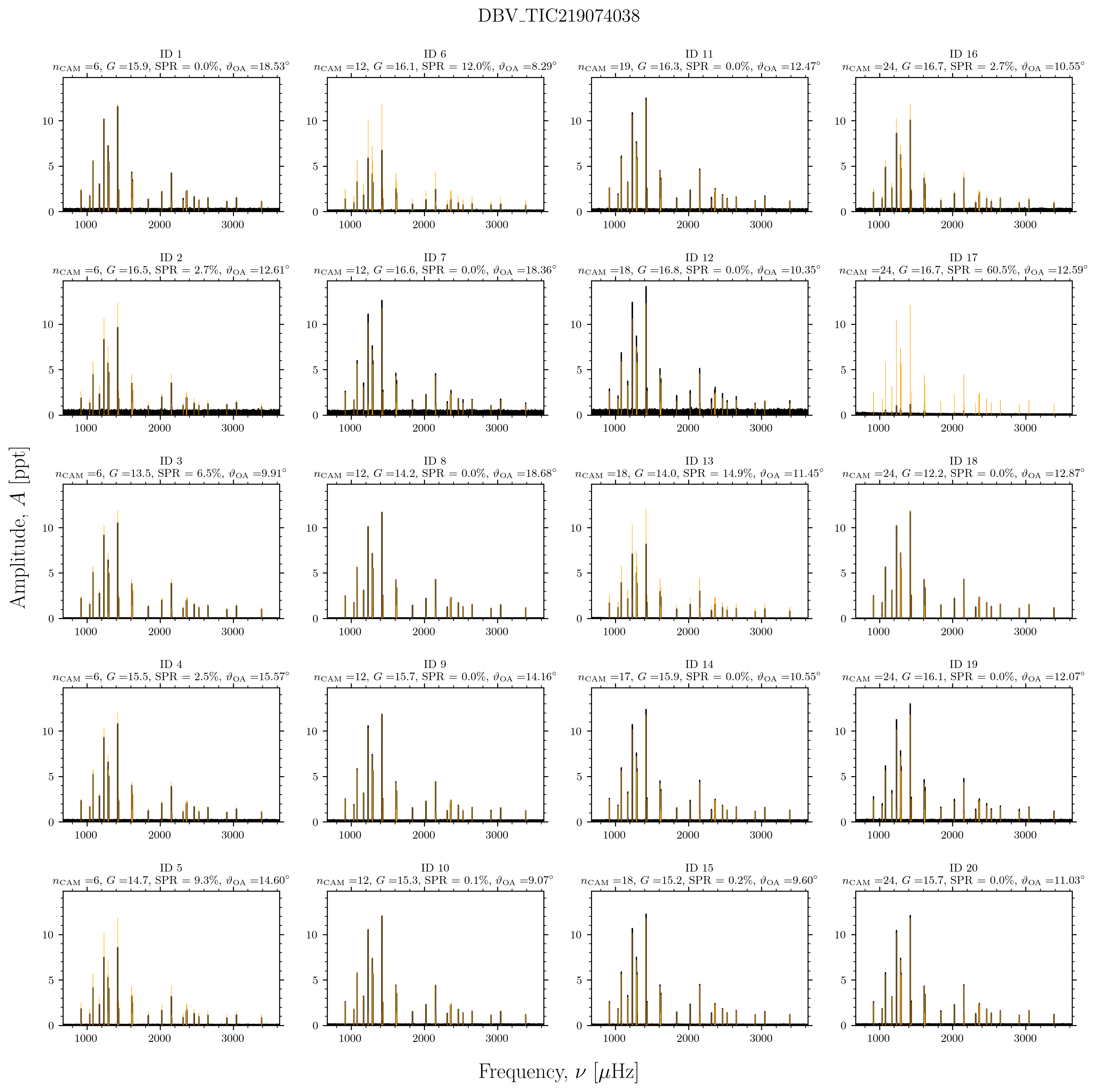}
\caption{Amplitude spectra of the DBV star TIC\,219074038 (GD358). Same description as Fig.~\ref{fig:amplitude_spectrum_DAV_TIC422526868}.} 
\label{fig:amplitude_spectrum_DBV_TIC219074038}
\end{figure*}

\begin{figure*}[h!]
\center
\includegraphics[width=1\columnwidth]{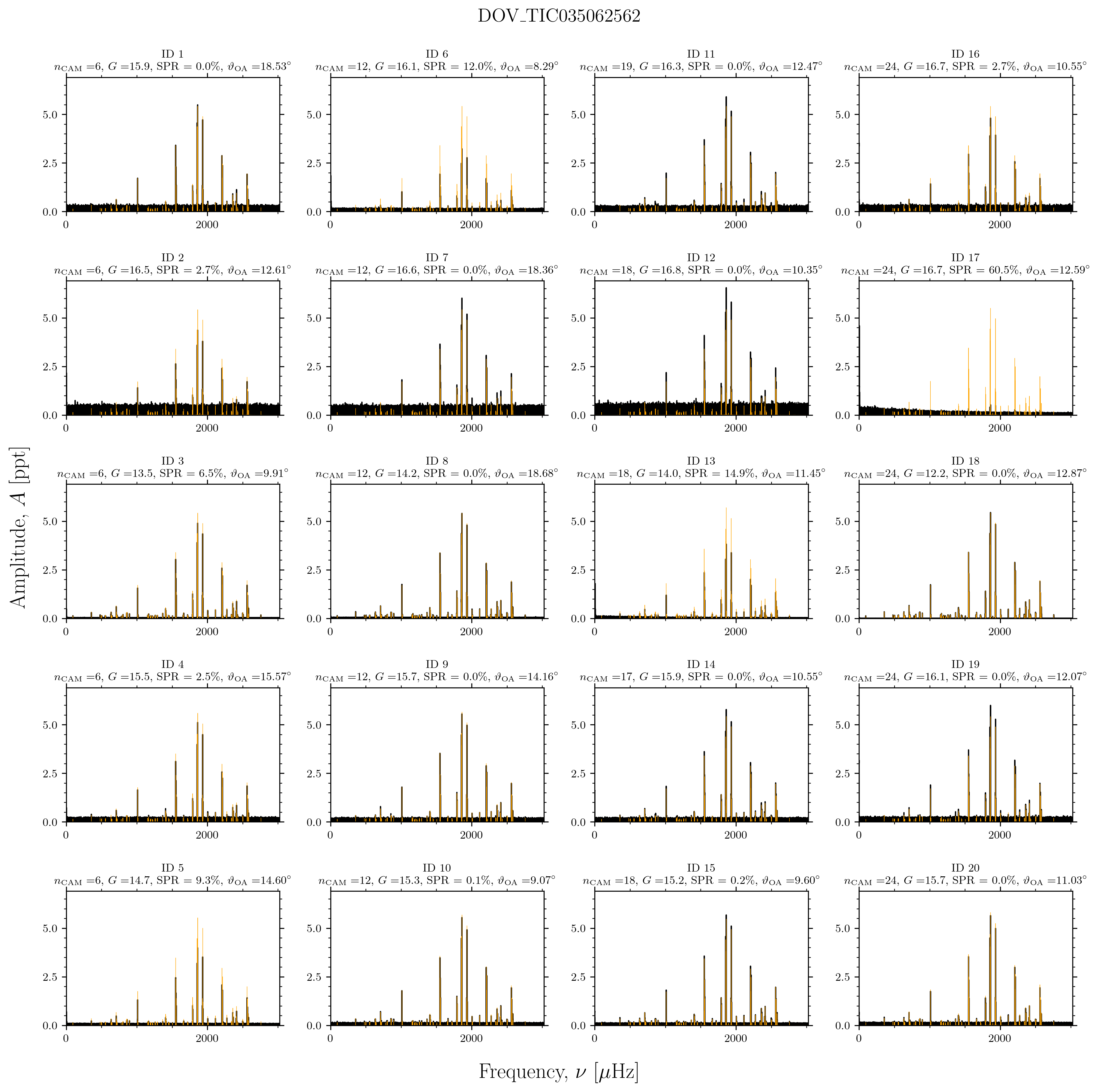}
\caption{Amplitude spectra of the GW vir star TIC\,035062562 (PG1159-035). Same description as for Fig.~\ref{fig:amplitude_spectrum_DAV_TIC422526868}.} 
\label{fig:amplitude_spectrum_DOV_TIC035062562}
\end{figure*}

\end{appendices}


\newpage
\bibliography{sn-bibliography}

\end{document}